# REGULATING DARK PATTERNS

MARTIN BRENNCKE[*]

*ABSTRACT*

Dark patterns have become increasingly pervasive in online choice architectures, encompassing practices like subscription traps, hiding information about fees, pre-selecting options by default, nagging, and drip pricing. Regulators around the world have started to express concerns that such practices are causing substantial consumer detriment. Efforts to effectively regulate dark patterns face the challenge that they often operate in the grey zone between legitimate persuasion techniques and clearly illegitimate methods of influencing consumer behavior such as coercion and deception.

This Article focuses on the legal response to dark patterns in the European Union, including the Digital Services Act, the Consumer Rights Directive, the Data Act, and the Consumer Credit Directive II. It provides the first comprehensive mapping of European Union laws expressly addressing dark patterns. The Article argues that these laws protect biased consumers and adopt autonomy as a normative lens to assess dark patterns. Consequently, regulating dark patterns in European Union law means regulating for autonomy. This normative lens is under-researched, and the existing literature has not yet produced a robust autonomy framework for regulating dark patterns. Developing such a framework is essential for any legal system aiming to regulate dark patterns to safeguard consumer autonomy, including the United States.

This Article addresses this gap in research with two principle contributions. First, it works out a specific conception of autonomous decision-making, rooted in the paradigm that providing consumers with information enables consumers to make an informed decision. This analysis challenges the dominant position in the literature, which holds that the information paradigm assumes that consumers are rational economic actors. Second, the Article offers a novel normative classification for dark patterns in online choice architectures. It develops a taxonomy encompassing six categories of autonomy violations, specifically tailored for the assessment and regulation of dark patterns that exploit consumer behavioral biases. These categories serve multiple purposes. They uncover and make explicit the autonomy violations addressed by existing European Union legislation. They delineate the contentious line between acceptable influences on consumer decision-making and autonomy violations that may warrant regulation in online choice architectures. They also provide policymakers in the EU and elsewhere with a framework when deliberating the regulation of other instances of dark patterns.

* Martin Brenncke, Ph.D., LL.M.; Senior Lecturer in Law at Aston Law School, Birmingham, United Kingdom. For valuable input, I am grateful to Rory van Loo.

TABLE OF CONTENTS

## INTRODUCTION

"Dark patterns" in online choice architectures are prevalent and increasingly used by businesses of all sizes.[1] Examples are default settings that maximize the collection of personal information, a button or website design that effectively hides fees, additional items automatically added to a consumer's online basket and website designs that make it easy to sign up to a service but very difficult to cancel or unsubscribe. Such practices can substantially influence consumer decision-making both at the individual and aggregate levels.[2] The growing use of artificial intelligence algorithms by businesses is said to offer significant possibilities to enhance the scale and efficiency of these and other dark patterns, thereby raising further concerns about consumer protection.[3]

These concerns have resonated with regulators globally. In 2020, the European Commission identified dark patterns as commercial practices that "disregard consumers' right to make an informed choice, abuse their behavioral biases, or distort their decision-making processes."[4] Since then, efforts to regulate dark patterns in the European Union (EU) have advanced significantly, for example with the Digital Services Act[5] and the ongoing digital fairness fitness check of existing consumer legislation.[6] The U.S. Federal Trade Commission has recognized the profound challenges dark patterns pose to consumers and is intensifying its efforts to regulate them.[7]

---

[1] *See, e.g.*, Linda Di Geronimo et al., *UI Dark Patterns and Where to Find Them: A Study on Mobile Applications and User Perception*, PROC. 2020 CONF. ON HUM. FACTORS COMPUTING SYS. 473:1–14 (2020); European Commission Press Release, *Consumer Protection: Manipulative Online Practices Found on 148 out of 399 Online Shops Screened* (Jan. 30, 2023), https://ec.europa.eu/commission/presscorner/detail/en/ip_23_418 [https://perma.cc/HKP3-EV58]; Francisco Lupiáñez -Villanueva et al., *Behavioural Study on Unfair Commercial Practices in the Digital Environment: Dark Patterns and Manipulative Personalisation. Final Report for the* EUROPEAN COMM'N at 6, 120 (Apr. 2022), https://op.europa.eu/en/publication-detail/-/publication/606365bc-d58b-11ec-a95f-01aa75ed71a1/language-en [https://perma.cc/S2MJ-XVXF]; Arunesh Mathur et al., *Dark Patterns at Scale: Findings from a Crawl of 11K Shopping Websites*, 3 PROC. ACM ON HUM.-COMPUTER INTERACTION 81:1-32 (2019); OECD, *Dark Commercial Patterns* 17–20 (OECD Digital Economy Papers, No. 336, 2022), https://www.oecd.org/digital/dark-commercial-patterns-44f5e846-en.htm. The term choice architecture refers to the environment in which consumers make decisions and how choices and information is presented to them.

[2] *See, e.g.*, OECD, *supra* note 1, at 21–23 (discussing various empirical studies evidencing the influence of dark patterns on consumer decision-making); Amit Zac et al., *Dark Patterns and Online Consumer Vulnerability* (Oxford Centre for Competition Law and Policy, Working Paper No. 55, 2023), https://papers.ssrn.com/sol3/papers.cfm?abstract_id=4547964 [https://perma.cc/3ZVB-GPXY].

[3] *See* Nathalie de Marcellis-Warin et al., *Artificial Intelligence and Consumer Manipulations: From Consumer's Counter Algorithms to Firm's Self-regulation Tools*, 2 AI & ETHICS 259–68 (2022); Sébastien Fassiaux, *Preserving Consumer Autonomy Through European Union Regulation of Artificial Intelligence: A Long-term Approach*, EUR. J. RISK REG. (forthcoming) (manuscript at 1, 7–9), https://doi.org/10.1017/err.2023.58 [https://perma.cc/3LKL-F43Q].

[4] EUROPEAN COMM'N, NEW CONSUMER AGENDA: STRENGTHENING CONSUMER RESILIENCE FOR SUSTAINABLE RECOVERY, at 10, COM (2020) 696 final (Nov. 13, 2020), https://commission.europa.eu/document/ac73e684-1e7f-4d36-a048-8f8a0b874448_en [https://perma.cc/KSH6-49FY].

[5] *See* Regulation (EU) 2022/2065 of the European Parliament and of the Council of 19 October 2022 on a Single Market for Digital Services and Amending Directive 2000/31/EC (Digital Services Act), art. 25, 2022 O.J. (L 277) 1 [hereinafter DSA]. *See, in detail, infra* Part I.C.

[6] *See* European Commission, *Digital Fairness – Fitness Check on EU Consumer Law*, EUROPEAN COMM'N, https://ec.europa.eu/info/law/better-regulation/have-your-say/initiatives/13413-Digital-fairness-fitness-check-on-EU-consumer-law_en [https://perma.cc/U9Z4-TTUS] (last visited Sept. 8, 2023).

[7] *See* FED.TRADE COMM'N, STAFF REPORT: BRINGING DARK PATTERNS TO LIGHT (Sept. 2022), https://www.ftc.gov/reports/bringing-dark-patterns-light [https://perma.cc/FFA6-W6RH].

Similarly, the OECD has expressed concerns that dark patterns may cause substantial consumer detriment.[8] A pivotal challenge in effectively reining in dark patterns is that they often operate in the grey zone between legitimate persuasion techniques and illegitimate methods of influencing consumer behavior.[9]

A clear normative foundation for regulating dark patterns in business-to-consumer (b2c) relationships is missing. The question of whether and the extent to which dark patterns necessitate regulatory intervention can be answered through various normative frameworks.[10] This Article contends that EU consumer law adopts autonomy as a normative lens to assess dark patterns. Dark patterns in online choice architectures undermine consumers' autonomy. This normative lens is under-researched, and the existing literature has not yet produced a well-developed autonomy framework for regulating dark patterns. One main hurdle for developing such a framework is that the meaning of autonomy has remained elusive.[11] An elusive notion of autonomy is unable to specify when dark patterns warrant regulation. This Article aims to bridge this gap in research by making two key contributions to the existing literature. First, the Article works out a specific conception of autonomous decision-making based on the information paradigm in EU consumer law. This analysis challenges the dominant viewpoint in the literature, which asserts that the information paradigm assumes that consumers are rational economic actors.[12] The Article posits that, despite the widespread critique that the information paradigm is ineffective in empowering consumers by providing them with information,[13] it actually holds the key for protecting consumers from dark patterns.[14] Empowerment

---

[8] See OECD, *supra* note 1, at 7.

[9] See Lupiáñez-Villanueva et al., *supra* note 1, at 6; Mario Martini & Christian Drews, *Making Choice Meaningful – Tackling Dark Patterns in Cookie and Consent Banners Through European Data Privacy Law*, 17–22 (2022), https://papers.ssrn.com/sol3/papers.cfm?abstract_id=4257979 [https://perma.cc/2TZR-EJ2S].

[10] See Arunesh Mathur et al., *What Makes a Dark Pattern . . . Dark?: Design Attributes, Normative Considerations, and Measurement Methods*, 3 PROC. 2021 CHI CONF. ACM ON HUM. FACTORS COMPUTING SYS. 360.-COMPUTER INTERACTION 81: 1–18 (2021) (distinguishing between individual welfare, collective welfare, individual autonomy, and other regulatory objectives).

[11] See Agnieszka Jabłonowska et al., *Consumer Law and Artificial Intelligence – Challenges to the EU Consumer Law and Policy Stemming From the Business' Use of Artificial Intelligence – Final Report of the ARTSY project* 12, 14 (EUI Working Paper LAW No. 11, Nov. 2018), https://cadmus.eui.eu/handle/1814/57484 [https://perma.cc/XZ66-V2FX]; Hans-Wolfgang Micklitz, *The General Clause on Unfair Practices*, in EUROPEAN FAIR TRADING LAW: THE UNFAIR COMMERCIAL PRACTICES DIRECTIVE 83, 104 (Geraint Howells et al. eds., 2006) (pointing out that the meaning of autonomy in the "entire European unfair trading law" seems largely incomprehensible).

[12] See *infra* Part II.D.

[13] See, e.g., Oren Bar-Gill & Omri Ben-Shahar, *Regulatory Techniques in Consumer Protection: A Critique of European Consumer Contract Law*, 50 COMMON MKT. L. REV. 109, 116–19 (2013); PAOLO SICILIANI ET AL., CONSUMER THEORIES OF HARM: AN ECONOMIC APPROACH TO CONSUMER LAW ENFORCEMENT AND POLICY MAKING 18–24 (2019). The empirical evidence is, however, inconclusive. See Rory van Loo, *Broadening Consumer Law: Competition, Protection, and Distribution*, 95 NOTRE DAME L. REV. 211, 245–48 (2019) (discussing empirical evidence which suggests that providing consumers with well-designed information improves choice in complex decision contexts and concluding that "the evidence suggests that well-crafted consumer law interventions can succeed in reducing consumer prices even in the face of adaptation by businesses"). *See also* Agnieszka Jabłonowska & Giacomo Tagiuri, *Rescuing Transparency in the Digital Economy: In Search of a Common Notion in EU Consumer and Data Protection Law*, Y.B. EUR. L. (forthcoming), https://doi.org/10.1093/yel/yead005 [https://perma.cc/X29W-QJ3C] (arguing that transparency can offer a meaningful frame for assisting consumers in the digital economy).

[14] *Cf.* Catalina Goanta & Cristiana Santos, *Dark Patterns Everything: An Update on a Regulatory Global Movement*, NETWORK L. REV. (Jan. 19, 2023),

and protection are deeply connected through their shared foundational conception of autonomy.

The Article's second key contribution lies in offering a normative classification for dark patterns in online choice architectures that exploit consumer behavioral biases. It develops a taxonomy comprising six categories of autonomy violations which are particularly relevant for evaluating and regulating dark patterns. These categories specify the contentious boundary between acceptable influences on consumer decision-making and autonomy violations that may warrant regulation in online choice architectures.[15] These categories not only concretize the existing EU laws governing dark patterns but also offer policymakers a framework when deliberating the regulation of other instances of dark patterns. While the literature does already feature numerous classifications of dark patterns, they grapple with two inherent limitations when viewed from the vantage point of EU consumer law.[16] They are either descriptive in nature, grounded in technical attributes of dark patterns, or they are normative but inadequately aligned with the provisions targeting dark patterns in EU consumer law.

While this Article focuses on the regulation of dark patterns in the EU, its theoretical and practical significance extends to other jurisdictions that employ autonomy as a normative lens to assess and regulate dark patterns. In the United States, for example, the California Privacy Rights Act (CPRA) is the first legislation explicitly addressing dark patterns.[17] The CPRA defines a dark pattern as: "[A] user interface designed or manipulated with the substantial effect of subverting or impairing user autonomy, decision-making, or choice, as further defined by regulation."[18] This definition underscores the CPRA's reliance on autonomy as a normative lens to assess and regulate dark patterns.[19] Other consumer privacy laws at the state level in Colorado, Connecticut, and Texas, for example, also expressly regulate the use of dark patterns because they subvert or impair user autonomy.[20] At the federal level, the Deceptive Experiences To Online Users Reduction (DETOUR) Act was introduced in Congress in April 2019.[21] Although the proposed Act does not explicitly mention dark patterns, it is aimed at prohibiting large online operators from using dark patterns in their user interfaces.[22] The Act would prohibit large online operators from employing any user interface "with the purpose or substantial effect of obscuring, subverting, or impairing user autonomy, or choice to obtain consent or user

---

https://www.networklawreview.org/digiconsumers-two/ [https://perma.cc/GC2F-S8SQ] (arguing that dark patterns should be discussed in the context of the information paradigm).

[15] *Cf.* Jennifer King & Adriana Stephan, *Regulating Privacy Dark Patterns in Practice: Drawing Inspiration From California Privacy Rights Act*, 5 GEO. L. TECH. REV. 250, 274 (2021); Jamie Luguri & Lior Jacob Strahilevitz, *Shining a Light on Dark Patterns*, 13 J. LEGAL ANALYSIS 43, 97–98, 102 (2021) (all emphasizing the challenge of delineating the boundary between illegitimate dark patterns and legitimate methods of persuasion).

[16] *See infra* Part III.C.

[17] California Privacy Rights Act, CAL. CIV. CODE § 1798.140 (2020) [hereinafter CPRA].

[18] *Id.* at §1798.140(l).

[19] *See* King & Stephan, *supra* note 15, at 265, 267.

[20] *See* Colorado Privacy Act, COLO. REV. STAT. § 6-1-1303 (2021); Connecticut Data Privacy Act, Public Act No. 22-15 (2022); Texas Data Privacy and Security Act, TEX. BUS. & COM. § 541.001 (2023).

[21] Deceptive Experiences to Online Users Reduction Act, S. 1084, 116th Cong. (2019) [hereinafter DETOUR Act].

[22] *See* Mark R. Warner, *Warner, Fischer Lead Bipartisan Reintroduction of Legislation to Ban Manipulative 'Dark Patterns'* (July 28, 2023), https://www.warner.senate.gov/public/index.cfm/2023/7/warner-fischer-lead-bipartisan-reintroduction-of-legislation-to-ban-manipulative-dark-patterns [https://perma.cc/Y6U5-745R].

data."[23] The reference to user autonomy in the proposed Act indicates that the legislation adopts autonomy as a normative lens to address dark patterns. The DETOUR Act was reintroduced in July 2023.[24] The connection between regulating dark patterns and safeguarding user autonomy in these laws exposes a research gap, as there is currently no well-developed autonomy framework for regulating dark patterns. This Article develops such a framework. The categories of autonomy violations within this framework may be of particular relevance for U.S. legislation, as they explain why specific dark pattern practices, currently under scrutiny by U.S. regulators such as the Federal Trade Commission,[25] violate autonomy and may warrant legal intervention.

The Article is structured as follows. Part I charts and elucidates the provisions in EU consumer law that either expressly protect consumers from dark pattern practices in online choice architectures or, equivalently, expressly protect consumers from commercial practices in online choice architectures that exploit consumer behavioral biases. Part II constitutes the Article's theoretical core. It specifies the process of decision-making which EU consumer law deems as autonomous by extracting a normative conception of autonomy from the information paradigm. Part III explicates when dark patterns that exploit consumer behavioral biases through the design of the online choice architecture violate autonomy. It develops six categories of autonomy violations in b2c relationships which hold particular pertinence in safeguarding biased consumers' autonomy in online choice architectures.

## I.    THE REGULATION OF DARK PATTERNS IN EU CONSUMER LAW

This Part comprehensively maps and explicates the provisions in EU consumer law that either expressly protect consumers from dark pattern practices in online choice architectures or, equivalently, expressly protect consumers from commercial practices in online choice architectures that exploit consumer behavioral biases.[26] A provision is considered to expressly regulate dark patterns when the statutory language, the recitals of the legislation, or the legislative materials specify that the provision is intended to regulate dark patterns.[27] This Part advances two arguments. First, these provisions (also) protect biased consumers. Second, the normative lens that these provisions adopt to assess dark patterns is autonomy.

---

[23] DETOUR Act § 3(a)(1).

[24] Deceptive Experiences to Online Users Reduction Act, S. 2708, 118th Cong (2023).

[25] *See* FED. TRADE COMM'N, *supra* note 7.

[26] This Article does not discuss Regulation (EU) 2022/1925 of the European Parliament and of the Council of 14 September 2022 on Contestable and Fair Markets in the Digital Sector and Amending Directives (EU) 2019/1937 and (EU) 2020/1828 (Digital Markets Act), 2022 O.J. (L 265) 1 [hereinafter DMA]. Even though the DMA contains provisions that are capable of reining in dark pattern practices of gatekeepers, such as the anti-circumvention provisions in DMA art. 13(4) and (6), the DMA complements competition law (*see* recitals 10–11 DMA) and has been characterised as a competition law instrument in the literature (*see* in detail Friso Bostoen, *Understanding the Digital Markets Act*, 68 ANTITRUST BULL. 263–306 (2023)). Hence, the DMA falls outside the scope of this Article.

[27] The same applies, *mutatis mutandis*, to provisions expressly protecting consumers from commercial practices in online choice architectures that exploit consumer behavioral biases.

### A.  THE MEANING OF DARK PATTERNS

Prior to commencing the analysis, it is imperative to gain more clarity about the meaning of the term dark patterns. The term lacks a commonly accepted definition in the literature.[28] Therefore, it seems apt to start with the definition adopted by EU law. The Digital Services Act (DSA) contains the first legal definition of dark patterns in EU law and defines dark patterns on online interfaces of online platforms as "practices that materially distort or impair, either on purpose or in effect, the ability of recipients of the service to make autonomous and informed choices or decisions."[29] Even though this Article uses the DSA's definition of dark patterns as a point of departure, two modifications appear warranted. First, the definition relies on the statutory language found in DSA art. 25(1), which seems to suggest that all dark patterns on online interfaces of online platforms are *per definitionem* prohibited by law. This Article, however, refrains from reserving the term solely for commercial practices that warrant legal prohibition. Instead, and in line with the existing literature,[30] it asks whether and when dark pattern practices warrant legal intervention.

Second, not all online choice architectures that violate the ability of consumers to make autonomous and informed decisions should be considered dark patterns. Otherwise, the term would amount to a vague general term covering a wide variety of types of influences.[31] For example, behavioural science regards coercion, which limits individuals' choice options, as a distinctly different means of influencing individuals compared to other forms of influence in online choice architectures which do not limit choice options.[32] In order to capture what is novel about dark pattern practices in online choice architectures, this Article limits these practices to online choice architectures that exploit behavioral biases.[33] This limitation is supported by the dominant view in scholarship according to which dark patterns (typically) exploit behavioral biases.[34] This limitation is also supported by the DSA's legislative materials. The incorporation of dark patterns into the Digital Services Act was a consequence of an amendment introduced by the European Parliament. The proposed amendment stated that dark patterns "typically exploit cognitive biases."[35]

---

[28] Colin M. Gray et al., *An Ontology of Dark Patterns: Foundations, Definitions, and a Structure for Transdisciplinary Action* 1 (2023), https://arxiv.org/pdf/2309.09640.pdf [https://perma.cc/Z2UL-P697].

[29] DSA, *supra* note 5, recital 67.

[30] *See, e.g.*, Luguri & Strahilevitz, *supra* note 15, at 82; Jan Trzaskowski, *Persuasion, Manipulation, Choice Architecture and 'Dark Patterns'* 27–28 (2023), https://papers.ssrn.com/sol3/papers.cfm?abstract_id=4491820 [https://perma.cc/58SW-CRK4].

[31] *Cf.* Trzaskowski, *id.* (arguing that the term dark pattern is not helpful from a legal perspective).

[32] See, e.g., Stuart Mills et al., *Dark Patterns and Sludge Audits: An Integrated Approach*, BEHAV. PUB. POL'Y (forthcoming) (manuscript at 2–3), https://doi.org/10.1017/bpp.2023.24 [https://perma.cc/H33Y-UY7Z].

[33] Even if one disagrees with this limitation of the term dark patterns, dark pattern practices are capable of capturing online choice architectures which exploit behavioral biases, and a key question for EU consumer law is to what extent biased consumers are protected from having their behavioral biases exploited by online choice architectures.

[34] *See, e.g.*, Luiza Jarovsky, *Dark Patterns in Personal Data Collection: Definition, Taxonomy and Lawfulness* 3 (2022), https://papers.ssrn.com/sol3/papers.cfm?abstract_id=4048582 [https://perma.cc/37QW-XB83]; King & Stephan, *supra* note 15, at 260; Martini & Drews, *supra* note 9, at 5; Mathur et al., *supra* note 1, at 6; Luguri & Strahilevitz, *supra* note 15, at 44; Szymon Osmola, *Neither Rules nor Standards: How to Regulate Dark Patterns* 9 (2023), https://papers.ssrn.com/sol3/papers.cfm?abstract_id=4515963 [https://perma.cc/2DDC-3FMV]; Sandeep Sharma & Ishita Sharma, *Dark Patterns in a Bright World: An Analysis of the Indian Consumer Legal Architecture*, 11 INT. J. CONSUMER L. & PRAC. 123, 140 (2023).

A consumer bias occurs when actual consumer judgment or decision-making systematically, rather than randomly, departs from a normative benchmark (from how a consumer ought to decide).[36] The dominant normative benchmark adopted in scholarship is rational choice theory (economic rationality), which refers to the optimal beliefs and choices assumed in the rational agent model dominant in neoclassical economics.[37] Defining bias relative to rational choice theory explains why coercive business practices fall outside the meaning of dark patterns. Coercion closes or narrows a consumer's choice options. It would not be ignored by a rational consumer and does not exploit consumer biases. For example, online choice architectures that alter or block off choice options affect the behavior of rational consumers. Examples of typical consumer biases are consumer choices that are influenced by default effects,[38] loss aversion, scarcity effects or the way in which information is presented to consumers (for example, salience or framing).[39] The field of behavioral economics has created a vast body of empirical research demonstrating that deviations from economic rationality are widespread in consumer decision-making and consistent in aggregate,[40] which makes them predictable and exploitable by commercial practices such as online choice architectures.

## B. *UNFAIR COMMERCIAL PRACTICES DIRECTIVE*

The starting point for considering dark patterns in EU consumer law is arguably the Unfair Commercial Practices Directive (UCPD),[41] which is the EU's flagship piece of legislation for regulating unfair commercial practices. The UCPD prohibits unfair commercial practices.[42] It does not expressly address dark patterns. Its text, recitals, and legislative materials do not refer to dark patterns or the exploitation of behavioral biases by commercial practices. Nonetheless, the European Commission is of the view that the broad and principle-based provisions in the UCPD are capable of challenging the unfairness of dark pattern practices.[43] Some scholars agree with this perspective.[44] The issue with this view is that it neglects the benchmark of the average consumer. When assessing the fairness of a commercial practice under the UCPD, the effect of the practice on consumer behavior is assessed from the perspective of the "average consumer."[45] The average consumer is said to be a rational economic actor,[46] and a rational economic actor is not biased. Hence, Rosca concludes that "[t]he average consumer . . . is not biased or susceptible to the exploitation of their cognitive biases."[47] This view severely limits the ability of the UCPD to capture commercial practices exploiting consumer behavioral biases. Even if one rejects this view, the question of whether the average consumer benchmark is sufficiently porous to incorporate behavioral findings about consumer heuristics (rules of thumb) and biases is an ongoing debate in EU consumer law scholarship.[48] The controversies surrounding this issue are one reason for the UCPD's limited effectiveness to curb dark patterns. As a consequence, the EU has decided to expressly regulate dark patterns in online choice architectures on online platforms in the Digital Services Act.

---

[35] *See* Digital Services Act recital 39a, as proposed by the European Parliament on 20 Jan. 2022, https://www.europarl.europa.eu/RegData/seance_pleniere/textes_adoptes/definitif/2022/01-20/0014/P9_TA(2022)0014_EN.pdf [https://perma.cc/QD6G-CVVT].

[36] *See* Martin Brenncke, *Reconceptualizing Behaviorally Informed Consumer Law and Policy*, 34 LOY. CONSUMER L. REV. 166, 168 (2022); JONATHAN ST.B.T. EVANS, HYPOTHETICAL THINKING: DUAL PROCESSES IN REASONING AND JUDGEMENT 2 (2007).

[37] *See, e.g.*, Brenncke, *supra* note 36, at 171–75 (giving an overview of the literature).

[38] A default in a choice context refers to the option pre-selected by the firm and that takes effect if the consumer does not make an active choice. Empirical evidence has shown that consumers are susceptible to accepting the default option in a choice context and that defaults can trigger behavior change in consumers. *See* N. Craig Smith et al., *Choice Without Awareness: Ethical and Policy Implications of Defaults*, 32 J. PUB. POL'Y & MARKETING 159, 160–61 (2013); Cass R. Sunstein, *Deciding by Default*, 162 U. PA. L. REV. 1, 11–17 (2013) (both reviewing the empirical evidence).

[39] Frames (framing effects) refer to different but formally equivalent descriptions of a decision problem that can give rise to different preferences and thus lead to different decisions. *See* Amos Tversky & Daniel Kahneman, *Rational Choice and the Framing of Decisions*, 59 J. BUS. S251 (1986).

[40] *See, e.g.*, Stefano DellaVigna, *Psychology and Economics: Evidence from the Field*, 47 J. ECON. LITERATURE 315–72 (2009); EYAL ZAMIR & DORON TEICHMAN, BEHAVIORAL LAW AND ECONOMICS ch. 15 (2018) (both discussing typical systematic deviations from the assumptions of rational choice theory in the consumer world, with reference to numerous empirical studies).

[41] Directive 2005/29/EC, of the European Parliament and of the Council of 11 May 2005 Concerning Unfair Business-to-consumer Commercial Practices in the Internal Market and Amending Council Directive 84/450/EEC, Directives 97/7/EC, 98/27/EC and 2002/65/EC of the European Parliament and of the Council and Regulation (EC) No 2006/2004 of the European Parliament and of the Council, 2005 O.J. (L 149) 22 [hereinafter UCPD].

[42] *Id.* at art. 5(1).

---

[43] *See* European Commission, *Commission Notice: Guidance on the Interpretation and Application of Directive 2005/29/EC of the European Parliament and of the Council Concerning Unfair Business-to-consumer Commercial Practices in the Internal Market*, 2021 O.J. (C 526) 4.2.7, 99–102 (Dec. 29, 2021), https://eur-lex.europa.eu/legal-content/EN/TXT/?uri=CELEX%3A52021XC1229%2805%29 [https://perma.cc/XEW5-NRLM].

[44] *See* Mark R. Leiser & Christiana Santos, *Dark Patterns, Enforcement, and the Emerging Digital Design Acquis – Manipulation Beneath the Interface* 21 (2023), https://papers.ssrn.com/sol3/papers.cfm?abstract_id=4431048 [https://perma.cc/YX44-253V]; Mark R. Leiser & Wen-Ting Yang, *Illuminating Manipulative Design: From 'Dark Patterns' to Information Asymmetry and the Repression of Free Choice Under the Unfair Commercial Practices Directive* (2022), https://papers.ssrn.com/sol3/papers.cfm?abstract_id=4418586 [https://perma.cc/AW76-C6K7]. *See*, however, Osmola, *supra* note 34, at 28–34 (arguing that the UCPD standards are not effective in tackling dark patterns).

[45] *See* UCPD, *supra* note 41, recital 18.

[46] *See* Rossella Incardona & Cristina Poncibò, *The Average Consumer, the Unfair Commercial Practices Directive, and the Cognitive Revolution*, 30 J. CONSUMER POL'Y 21, 30–31 (2007); Vanessa Mak, *The Consumer in European Regulatory Private Law*, in THE IMAGE(S) OF THE CONSUMER IN EU LAW 381, 386–88 (Dorota Leczykiewicz & Stephen Weatherill eds., 2016); Christine Riefa & Harriet Gamper, *Economic Theory and Consumer Vulnerability: Exploring an Uneasy Relationship*, in VULNERABLE CONSUMERS AND THE LAW: CONSUMER PROTECTION AND ACCESS TO JUSTICE 18, 20 (Christine Riefa & Séverine Saintier eds., 2021); Cătălin Gabriel Stănescu, *The Responsible Consumer in the Digital Age*, 24 TILBURG L. REV. 49, 53 (2019).

[47] Constanta Rosca, *Destination 'Dark Patterns': On the EU (Digital) Legislative Train and Line-drawing* (Apr. 13, 2023), https://www.maastrichtuniversity.nl/blog/2023/04/destination-%E2%80%98dark-patterns%E2%80%99-eu-digital-legislative-train-and-line-drawing [https://perma.cc/X4VF-5FQB]. *See also* Marie Jull Sørensen et al., *Response of the European Law Institute to the European Commission's Public Consultation on Digital Fairness – Fitness Check on EU Consumer Law* 7–8 (Apr. 20, 2023), https://www.europeanlawinstitute.eu/fileadmin/user_upload/p_eli/Publications/Response_of_the_ELI_to_the_European_Commission_s_Public_Consultation_on_Digital_Fairness_.pdf [https://perma.cc/UJ9Q-Z5BK] (noting that "the whole idea of dark patterns sits uneasily with the UCPD benchmark of an average consumer, as it has traditionally been understood" and that "[i]deally, a UCPD average consumer does not have cognitive biases"); Eliza Mik, *The Erosion of Autonomy in Online Consumer Transactions*, 8 L., INNOVATION & TECH. 1, 33–34 (2016) ("the concept of 'average consumer' not only ignores the manner in which consumers act in practice but also disregards the very findings in behavioural economics and cognitive science that online businesses rely on to influence consumer behaviour. If certain cognitive biases are known to affect *most* market participants and if they are commonly exploited it is dangerous to pretend they do not exist.").

[48] *See, e.g.*, Jason Cohen, *Bringing Down the Average: The Case for a "Less Sophisticated" Reasonableness Standard in US and EU Consumer Law*, 32 LOY. CONSUMER L. REV. 1, 33–36, 40–44 (2019).

## C. *DIGITAL SERVICES ACT*

DSA art. 25(1) prohibits providers of online platforms such as social media, content-sharing websites (for example, Twitter, Facebook, LinkedIn, Instagram, TikTok, Youtube), and online marketplaces (for example, Amazon Store) to design, organize or operate their "online interfaces"[49] "in a way that deceives or manipulates the recipients of their service or in a way that otherwise materially distorts or impairs the ability of the recipients of their service to make free and informed decisions."[50] DSA art. 25 also empowers the European Commission to issue guidelines on how the prohibition applies to specific commercial practices.[51] The prohibition, which will apply from February 2024 in the European Union, predominantly targets dark patterns.[52] The prohibition protects all service recipients of providers of online platforms, which includes consumers and business users.[53] The legislative materials[54] and the reference to "bias"[55] in the DSA confirm that DSA art. 25(1) (also) protects biased consumers.[56] This is further evidenced by the fact that the DSA acknowledges that "presenting choices in a non-neutral manner, such as giving more prominence to certain choices through visual, auditory, or other components" affects consumer decision-making.[57] The salient or non-salient presentation of information or choice options can affect how consumers evaluate this information or option in their decision-making process, which means that their choice is context-dependent and, therefore, biased relative to rational choice theory.[58]

Turning our attention to the policy objective pursued by DSA art. 25(1), the definition of dark patterns in the DSA makes explicit the link between protecting service recipients' ability to make free and informed decisions and protecting their autonomy.[59] The DSA also distinguishes between "decision-making and choice,"[60] thus acknowledging that the protective scope of DSA art. 25(1) is not limited to one's choice options but also includes one's decision-making. This aligns with autonomy theory, which recognizes the independence of one's decision-making and of one's

---

[49] "[O]nline interface" means any software, including a website or a part thereof, and applications, including mobile applications (DSA, *supra* note 5, art. 3(m)).

[50] *Id.* at art. 25(1).

[51] *See id.* at art. 25(3), which also lists three practices, for example the practice of "nagging" in *id.* at art. 25(3)(b).

[52] *See id.* at recital 67. Note that DSA art. 25(1) is not limited to dark patterns but captures all structures, designs or functionalities of an online interface that materially distort or impair "the ability of the recipients of their service to make free and informed decisions".

[53] *See id.* at recital 2. According to DSA art. 1, consumer protection is one of the aims of the DSA.

[54] *See id.* at recital 39a, as proposed by the European Parliament on 20 Jan. 2022, https://www.europarl.europa.eu/RegData/seance_pleniere/textes_adoptes/definitif/2022/01-20/0014/P9_TA(2022)0014_EN.pdf [https://perma.cc/7UR2-5A5H] ("certain practices typically exploit cognitive biases").

[55] *Id.* at recital 67.

[56] Note that the scope of DSA, *id.* at art. 25(1), is not limited to dark patterns and the protection of biased consumers but extends to protecting all consumers, including rational consumers, whose ability to make free and informed decisions is materially distorted or impaired by online choice architecures. For rational consumers, however, the UCPD, *supra* note 41, will often be lex specialis (*cf.* DSA art. 25(2)).

[57] *Id.* at recital 67. *See also id.* at art. 25(3)(a).

[58] *See, e.g.,* Pedro Bordalo et al., *Salience and Consumer Choice,* 121 J. POL. ECON. 803–43 (2013); Matthew D. Hilchey et al., *Does the Visual Salience of Credit Card Features Affect Choice?,* 7 BEHAV. PUB. POL'Y 291–308 (2023).

[59] *See* DSA, *supra* note 5, recital 67.

[60] *Id.*

choice.[61] Hence, the words "free and informed decisions" in DSA art. 25(1) incorporate both dimensions of autonomy. It is worth emphasizing in this context that choice architectures that exploit consumer biases preserve choice options, but they interfere with consumers' decision-making, for example by changing how consumers understand their options. To conclude, the Digital Services Act adopts autonomy as a normative lens to assess dark patterns.

### D. DIRECTIVE ON FINANCIAL SERVICES CONTRACTS CONCLUDED AT A DISTANCE

One of the objectives of the Directive on Financial Services Contracts Concluded at a Distance[62] is to prevent traders, when concluding financial services contracts at a distance, from using dark patterns in their online interfaces. The Directive aims to achieve this objective by inserting a new art. 16e into the Consumer Rights Directive (CRD).[63] This new provision closely mirrors both the text and material scope of DSA art. 25, reflecting that it was drafted against the background of the Digital Services Act.[64] Consequently, CRD art. 16e should be interpreted in the same way as DSA art. 25 wherever their statutory language is identical.

The Directive on Financial Services Contracts Concluded at a Distance largely features the same definition of dark patterns as that found in the Digital Services Act.[65] The Directive also makes explicit the link between protecting consumers' ability to make free and informed decisions and protecting their autonomy.[66] Furthermore, the specific commercial practices listed in DSA art. 25(3) are replicated in CRD art. 16e with nearly identical wording. As explained in the previous section regarding DSA art. 25(1), these factors demonstrate that CRD art. 16e protects biased consumers. This is also supported by the legislative materials, as the following reasoning shows. The Impact Assessment accompanying the European Commission's proposal for the Directive on Financial Services Contracts Concluded at a Distance contains the most explicit discussion of commercial practices that exploit consumer behavioral biases in EU legislative materials to-date.[67]

---

[61] *See* John Christman, *Autonomy in Moral and Political Philosophy, in* STANFORD ENCYCLOPEDIA OF PHILOSOPHY 1.2 (Edward N. Zalta ed., 2015) (Jun. 29, 2020), https://plato.stanford.edu/entries/autonomy-moral/ [https://perma.cc/LH7V-H295] ("the independence of one's deliberation and choice from manipulation by others"); JOSEPH RAZ, THE MORALITY OF FREEDOM 377–78 (1986) (distinguishing between limitations of autonomy that diminish a person's options or pervert the way a person reaches decisions, forms preferences, or adopts goals).

[62] Directive (EU) 2023/2673 of the European Parliament and of the Council of 22 November 2023 Amending Directive 2011/83/EU as Regards Financial Services Contracts Concluded at a Distance and Repealing Directive 2002/65/EC, 2023 O.J. (L 21) 1 [hereinafter Directive On Financial Services Contracts Concluded at a Distance].

[63] *Id.* at art. 1(4). *See also* Directive 2011/83/EU of the European Parliament and of the Council of 25 October 2011 on Consumer Rights, Amending Council Directive 93/13/EEC and Directive 1999/44/EC of the European Parliament and of the Council and Repealing Council Directive 85/577/EEC and Directive 97/7/EC of the European Parliament and of the Council, 2013 O.J. (L 176) 338 [hereinafter CRD].

[64] *See also* the largely identical DSA, *supra* note 5, recital 67; Directive on Financial Services Contracts Concluded at a Distance, *supra* note 62, recital 41.

[65] *See* Directive on Financial Services Contracts Concluded at a Distance, *supra* note 62, recital 41.

[66] *Id.*

[67] *See* European Commission, *Impact Assessment Report Accompanying the Proposal for a Directive of the European Parliament and of the Council Amending Directive 2011/83/EU Concerning*

According to the Impact Assessment, the increasing digitalization of financial services has led to new market practices that exploit consumer behavioral biases.[68] The Impact Assessment explicitly identifies these market practices as a problem driver that has led to consumer detriment, which has remained unaddressed by older EU legislation.[69] To alleviate this consumer detriment, the European Commission proposed CRD art. 16e, which "aims to ensure that traders [when concluding financial services contracts at a distance] do not benefit from consumer biases. In this light, they are prohibited from setting up their online interfaces in a way which can distort or impair the consumers' ability to make a free, autonomous and informed decision or choice."[70] To conclude, CRD art. 16e protects biased consumers and adopts autonomy as a normative lens to assess dark patterns.

### E. *DATA ACT*

The Data Act[71] regulates the sharing of data, obtained from or generated by the use of a product or related service, between users, data holders, and third parties. The Data Act specifically targets data generated by internet-connected devices (Internet of Things) and is expected to promote competition in aftermarkets for Internet-of-Things products.[72] For example, users of connected products are granted a right to obtain access to data generated by the use of the product from the data holder (for example, the product's manufacturer) where this data is not already directly accessible to the user.[73] Users can then share this data with repair or service providers (third parties), fostering competition in the aftermarket. When users exercise their right to obtain access to data, "third parties or data holders should not rely on so-called 'dark patterns' in designing their digital interfaces."[74] Data Act art. 4(4) formulates the prohibition of dark patterns in the Data Act as follows:

> Data holders shall not make the exercise of choices or rights under this Article by the user unduly difficult, including by offering choices to the user in a non-neutral manner or by subverting or impairing the autonomy, decision-making or choices of the user via the structure, design, function or manner of operation of a user digital interface or a part thereof.[75]

A very similar prohibition of dark patterns also exists for third parties who receive data at the request of a user.[76] This applies, for example, when third parties present choices to users which relate to the purpose of data processing or the deletion of data that is no longer necessary for the agreed purpose.[77]

Even though the Data Act is not consumer law, it complements consumer law.[78] In the digital economy, data sharing affects individuals not only in their capacity as users but also in their capacity as consumers. Consequently, the prohibition of dark patterns in the Data Act protects consumer users of connected products and related services. This is clear from the title of Chapter II of the Data Act, which implies that the prohibitions of dark patterns in the Data Act apply to b2c data sharing, that is, b2c relationships. The prohibitions of dark patterns for data holders and third parties (also) protect biased consumers.[79] This is not only evidenced by the reference to "bias" in the Data Act,[80] but also by the statutory words. That is because the prohibitions of dark patterns for data holders and third parties in the Data Act acknowledge that salience effects, which are ignored by a rational economic agent, can influence user decision-making.[81]

The Data Act also adopts a definition of dark patterns. "Dark patterns are design techniques that push or deceive consumers into decisions that have negative consequences for them."[82] Regrettably, this definition differs from the definition of dark patterns in the DSA. It is unclear whether this lack of coherence has much relevance.[83] One obvious distinction between both definitions is that the autonomy violation, which is characteristic for the definition of dark patterns in the Digital Services Act, finds no parallel in the definition of dark patterns in the Data Act. Nevertheless, both the recitals and the statutory text of the Data Act explicitly make reference to users' autonomy as a protected good.[84] The prohibitions of dark patterns in the Data Act protect users from digital interfaces that subvert or impair their autonomy, essentially embracing autonomy as a normative lens for assessing dark patterns. The Data Act also distinguishes between "decision-making or choices of the user."[85] This echoes the same distinction drawn in the DSA, and this differentiation acknowledges that autonomy protects users' choice options and their decision-making.

### F.   CONSUMER RIGHTS DIRECTIVE AND CONSUMER CREDIT DIRECTIVE II

The Digital Services Act, the amended Consumer Rights Directive, and the Data Act may be the only EU legislative acts so far that expressly use the term dark patterns, but they are not the only EU legislative acts that expressly protect consumers from online choice architectures that exploit their behavioral biases. CRD art. 22 grants the consumer a right to reimbursement of any payment in addition to the remuneration for the trader's main contractual obligation for which the consumer's consent was inferred "by using default options which the consumer is required to reject in order to avoid the additional payment."[86] An illustrative instance of such a commercial practice is the pre-ticking of a checkbox for purchasing travel insurance when the consumer buys a flight ticket online. Consumers must actively untick the checkbox if they wish to decline the travel insurance. The mention of default options in CRD art. 22 distinctly points to default effects as the rationale underpinning this provision.

---

[76] *See id.* at art. 6(2)(a).

[77] *Cf. id.* at art. 6(1).

[78] *See id.* at art. 1(9) and recital 9.

[79] Note that the scope of *id.* at arts. 4(4) and 6(2)(a) is not limited to dark patterns and the protection of biased consumers but extends to protecting all consumers, including rational consumers, whose autonomy is subverted or impaired by the design of the user interface.

[80] *Id.* at recital 38.

[81] *See id.* at arts. 4(4) and 6(2)(a) ("[I]ncluding by offering choices to the user in a non-neutral manner. . . .").

[82] *Id.* at recital 38.

[83] Despite the differences in the definition of dark patterns, *id.* is otherwise clearly aligned with DSA, *supra* note 5, recital 67.

[84] *See* Data Act, *supra* note 71, recital 38 and arts. 4(4) and 6(2)(a).

[85] *Id.* at arts. 4(4) and 6(2)(a).

This is confirmed by the Consumer Credit Directive II (CCD II), art. 15.[87] This provision prohibits inferring an agreement for the conclusion of any consumer credit or the purchase of ancillary services through the presentation of default options like pre-ticked boxes. The Impact Assessment that accompanies the European Commission's proposal for the Consumer Credit Directive II considers the regulation of pre-ticked boxes as a strategy for addressing credit providers' exploitation of consumer biases.[88] The Impact Assessment also acknowledges consumer biases as a driving force behind the default effect linked with pre-ticked boxes.[89] Hence, preventing the exploitation of consumers' behavioral biases serves as the rationale for regulating default options such as pre-ticked boxes in (online) choice environments.[90] The regulation of default options in the CRD and the CCD II protect biased consumers.

Both CRD art. 22 and CCD II art. 15 also aim to protect consumer autonomy as the following argumentation shows. The former provision requires traders to obtain consumers' "express consent" for any payment in addition to the remuneration agreed upon for the trader's main contractual obligation. The latter provision requires creditors and credit intermediaries to obtain consumers' "freely given, specific, informed and unambiguous" consent for the conclusion of any consumer credit or the purchase of ancillary services presented through boxes. Default options such as pre-ticked boxes fail to satisfy these consent requirements. They violate consumer autonomy, because respect for autonomy functions as a predominant rationale for informed consent requirements.[91]

---

[86] CRD, *supra* note 63, art. 22.
[87] Directive (EU) 2023/2225 of the *European* Parliament *and of the Council* of 18 October 2023 on Credit Agreements for Consumers and Repealing Directive 2008/48/EC, 2023 O.J. (L 67) 1 at 15 [hereinafter CCD II].
[88] *See* European *Commission, Staff Working Document Impact Assessment Report Accompanying the Proposal for a Directive of the European Parliament and of the Council on Consumer Credits*, at 64, SWD (2021) 170 final (June 30, 2021), https://eur-lex.europa.eu/legal-content/EN/TXT/?uri=SWD%3A2021%3A0170%3AFIN [https://perma.cc/9LVX-JJ8V].
[89] See *id.* at 15–16.
[90] *See also* European Consumer Organisation (BEUC), *"Dark Patterns" and the EU Consumer Law Aquis: Recommendations for Better Enforcement and Reform*, The Consumer Voice in Eur. 10 (Feb. 7, 2022), https://www.beuc.eu/sites/default/files/publications/beuc-x-2022-013_dark_patterns_paper.pdf [https://perma.cc/5FJF-UZSJ] (stating that "the very essence of the prohibition [in CRD, *supra* note 63, art. 52] was to prevent traders from taking advantage of consumers' status quo biases").
[91] *See, e.g.*, Ruth R. Faden & Tom L. Beauchamp, A History and Theory of Informed Consent ch. 7, 8 (1986); Lucie White, *Understanding the Relationship Between Autonomy and Informed Consent: A Response to Taylor*, 47 J. Value Inquiry 483–91 (2013).

G.  *REGULATION ON KEY INFORMATION DOCUMENTS FOR PACKAGED RETAIL*
   *INVESTMENT AND INSURANCE-BASED INVESTMENT PRODUCTS*

The Regulation on Key Information Documents for Packaged Retail Investment and Insurance-based Investment Products (PRIIPs Regulation)[92] is another legislative act that does not explicitly use the term dark patterns but expressly protects consumers from online choice architectures that exploit their behavioral biases.[93] The Regulation introduces a key information document (KID), which financial consumers must receive prior to their investment in a packaged retail investment or insurance-based investment product (PRIIP). The Regulation ensures uniformity in KID content and presentation through highly prescriptive rules, specifying elements such as the exact wording of headings, the sequencing of information, or a comprehension alert and its specific wording.[94]

The KID serves as an instance of state-driven informational choice architecture, applicable both online and offline. Its objective is to allow financial consumers to better comprehend complex investment products and better compare different investment options.[95] By fostering greater understandability and comparability of information, the KID aims to enable financial consumers to make better informed investment decisions.[96] To achieve these objectives, the KID prioritizes prescriptive rules over high-level principles. The European Commission's Impact Assessment suggests that if KIDs were based solely on high-level principles like accuracy, fairness, and clarity, the presentation of information in KIDs could be "gamed (enabling subtle investor biases to be exploited)."[97] For instance, the Impact Assessment offers the example that "subtle juxtapositions and hierarchies of information (for instance, placing cost information on a back page) can have strong impacts as to how salient information is taken to be for retail investors."[98] Through regulating the presentation of information in the KID in detail, the PRIIPs Regulation asserts control over the impact of salience on consumer decision-making.

The KID can therefore be seen as a regulatory instrument safeguarding consumers susceptible to biases, and its purpose encompasses a dual, enabling and protective, rationale. It proactively shapes the informational choice architecture in order to (i) enable and facilitate consumer comprehension and decision-making and (ii) protect consumers from harmful firm conduct that exploits consumer biases through informational choice architecture. In relation to the latter purpose, the PRIIPs Regulation implies that the exploitation of consumer behavioral biases in KIDs would prevent financial consumers from making better informed investment decisions. This links the KID's protective rationale to autonomy. Autonomy serves as the normative benchmark for assessing choice architectures that exploit consumer behavioral biases in KIDs. Even though this is not obvious from the face of the Regulation, the PRIIPs KID is an example of the EU's predominant tool of consumer protection, the information paradigm. This Article argues in the next Part that the information paradigm enables and empowers consumers to make autonomous decisions. Hence, the PRIIPs KID facilitates and protects financial consumers' autonomy.

---

[92] Regulation (EU) No 1286/2014, of the European Parliament and of the Council of 26 November 2014 on Key Information Documents for Packaged Retail and Insurance-based Investment Products (PRIIPs), Dec. 9, 2014, 2014 O.J. (L 352) 1 [hereinafter PRIIPs Regulation].

[93] *Id.* at recitals 17 and 27 offer clear confirmation that retail investors are consumers. In this Article, I use the phrase "financial consumer" as a synonymous term for "retail investor" in the PRIIPs Regulation.

[94] *See id.* at art. 8.

[95] *See id.* at art. 1, recitals 1, 36.

[96] *See id.* at recitals 15, 26.

[97] European Commission, *Impact Assessment Accompanying the Proposal for a Regulation of the European Parliament and of the Council on Key Information Documents for Investment Products*, at 36, SWD (2012) 187 final (July 10, 2012), https://data.consilium.europa.eu/doc/document/ST-12402-2012-ADD-1/en/pdf [https://perma.cc/7GPJ-U6H4].

[98] *Id.*

## H. Conclusion

This Part has shown that multiple legislative acts in the EU either expressly protect consumers from dark pattern practices in online choice architectures or, equivalently, expressly protect consumers from commercial practices in online choice architectures that exploit consumer behavioral biases. The key provisions are DSA art. 25, CRD art. 16e, Data Act arts. 4(4) and 6(2)(a), CRD art. 22, CCD II art. 15 and the Key Information Document in the PRIIPs Regulation. These provisions protect biased consumers, whose behavior deviates from rational choice theory. Biased consumers require protection from having their biases exploited by commercial practices.[99]

This prompts the inquiry into the normative theory capable of elucidating why the exploitation of consumer biases in online choice architectures is wrong and warrants legal intervention. The analysis in this Part has shown that the normative theory adopted by EU consumer law is autonomy. The provisions scrutinized in this Part regulate specific cases of autonomy violations. What is problematic is that these provisions and legislative acts do not define or specify the meaning of autonomy. The existing literature has not yet closed this gap either. It has not yet produced a well-developed conception of autonomous decision-making for the purposes of EU consumer law. Such a conception is needed in order to specify the autonomy violation which underlies the provisions governing dark patterns in EU consumer law. Uncovering these autonomy violations and making them explicit helps interpreting these provisions, applying them in borderline cases, and making them more effective tools of consumer protection. This is particularly relevant for the prohibitions of dark patterns in the DSA, the CRD, and the Data Act, as these provisions adopt various vague legal terms. Martini and Drews, for example, have pointed out that the vagueness in the criteria employed by DSA art. 25(1) creates legal uncertainty and undermines the effectiveness of the provision to curb dark patterns in practice.[100]

---

[99] Zac et al., *supra* note 2, provide a detailed theoretical discussion and empirical analysis of whether biased consumers, who are exposed to dark pattern practices, qualify as vulnerable consumers. They test the susceptibility of different user groups to a range of dark patterns and find that (i) all user groups can be adversely affected by dark patterns and (ii) dark patterns are effective in influencing user decision-making regardless of users' income, education, or age. *Cf.* Martin Brenncke, *A Theory of Exploitation for Consumer Law: Online Choice Architectures, Dark Patterns, and Autonomy Violations*, J. Consumer Pol'y (forthcoming) (manuscript at 8–11), https://doi.org/10.1007/s10603-023-09554-7 [https://perma.cc/8ZDD-6CLN] (analyzing when online choice architectures exploit biased consumers and arguing that vulnerability is not a necessary condition for an exploitation claim in consumer law).

[100] *See* Martini & Drews, *supra* note 9, at 28–29. *See also* Osmola, *supra* note 34, at 41 (assessing that DSA art. 25 "adds more uncertainty with regard to norms applicable to particular circumstances and thus causes more harm than good in combating dark patterns").

The next Part of this Article works out a specific conception of autonomous decision-making for EU consumer law, and Part III of this Article uncovers and makes explicit the autonomy violations targeted by the provisions regulating dark patterns in EU consumer law. This will specify the contentious line between acceptable influences on consumer decision-making and autonomy violations in online choice architectures. Not only is this specification important for understanding the prohibitions of dark patterns in the DSA, the CRD, and the Data Act, but also for future regulatory activity targeting dark patterns. Such legislative activity is certainly on the horizon.[101]

## II.   THE MEANING OF AUTONOMY IN EU CONSUMER LAW

This Part of the Article contributes to the literature by (i) elucidating how the information paradigm in EU consumer law supports consumer autonomy, and (ii) extracting a normative conception of autonomous decision-making from the information paradigm. Resorting to the information paradigm is justified since (i) the analysis of the provisions governing dark patterns in the previous Part of this Article was unable to specify the meaning of autonomy and (ii) the information paradigm is the most important tool of consumer policy in the EU. This Part shows that the information paradigm and the provisions governing dark patterns operate with the same underlying conception of autonomy. While the purpose of the information paradigm can be conceptualised as enabling consumers to make model autonomous decisions, the purpose of the dark pattern provisions in EU consumer law is to protect consumers' ability to make model autonomous decisions. Hence, the meaning of autonomy enshrined in the information paradigm forms the benchmark for assessing whether dark patterns in b2c relationships violate consumer autonomy. The analysis in this Part also challenges the dominant position in the literature, which asserts that the information paradigm in EU consumer law assumes that consumers are rational economic actors.[102] Contrary to Siciliani et al., the information paradigm in EU consumer law is not "borne out of an over-zealous embrace of neoclassical economics."[103] Instead, it can have an autonomy-based justification.

### A.   AUTONOMY AS A PHILOSOPHICAL AND LEGAL CONCEPT

Before delving into this justification, a note on methodology is warranted. Autonomy refers to self-governance.[104] Apart from this common ground, the meaning of autonomy in philosophical scholarship is the subject of considerable debate.[105] This is a significant hurdle for using the philosophical discourse as a starting point for determining the meaning of autonomy for the purposes of EU consumer law or legal purposes in general. Ashcroft rightly points out that when philosophical concepts such as autonomy are used in the law, they should be interpreted in their specific context of use.[106] Following Ashcroft, this Article develops a technical meaning of autonomy for the specific context of EU consumer law. This approach diverges from other perspectives in the literature.[107] While those perspectives posit that autonomy can serve as a theoretical foundation for behaviorally informed consumer law, they neither derive their normative conception of autonomy from EU consumer law nor examine its compatibility with EU consumer law.

---

[101] See the ongoing digital fairness fitness check in EU consumer law, assessing whether existing EU consumer law ensures a high level of protection in the digital environment (see European Commission, *Digital Fairness*, *supra* note 6).

[102] See, e.g., Incardona & Poncibò, *supra* note 46, at 31–33; Mak, *supra* note 46, at 386–88; Andreas Oehler & Stefan Wendt, *Good Consumer Information: The Information Paradigm at its (Dead) End?*, 40 J. CONSUMER POL'Y 179, 181 (2017); Stănescu, *supra* note 46, at 53.

[103] SICILIANI ET AL., *supra* note 13, at 21.

[104] See, e.g., Tom L. Beauchamp, *Autonomy and Consent, in* THE ETHICS OF CONSENT: THEORY AND PRACTICE 55, 61 (Franklin G. Miller & Alan Wertheimer eds., 2009); BEN COLBURN, AUTONOMY AND LIBERALISM 4 (2010).

While the analysis in this Article is primarily legalistic in nature, rather than philosophical, it will be conducted within the framework of the core conditions of personal autonomy that are largely accepted in philosophical discourse.[108] As acknowledged even by Ashcroft, the extra-legal philosophical framework helps structuring, contextualising, and clarifying the legal analysis.[109] According to Christman, personal autonomy refers to the "capacity to be one's own person, to live one's life according to reasons and motives that are taken as one's own and not the product of manipulative or distorting external forces, to be in this way independent."[110] This definition of personal autonomy contains two core conditions: self-determination and procedural independence.[111]

Whereas procedural independence refers to one's relationship with others, self-determination is an internal condition of autonomous choice and refers to one's relationship with oneself.[112] For the purposes of consumer law, self-determination can be linked to consumer empowerment and procedural independence can be linked to consumer protection. Procedural independence means that decision-making must be independent from distorting external influences such as coercion and manipulation.[113] It is relatively uncontroversial that not every successful external influence on the decision-making process is distorting and a violation of autonomy. Autonomy is not purely individualistic but relational and in principle compatible with external factors that influence individuals' decision-making.[114] As Sneddon explains, self-governance does not require that an individual is independent from significant contributions from the social environment.[115] The difficulty lies in distinguishing between distorting, autonomy-violating external influences and other external influences that are compatible with autonomy.

---

[105] See, e.g., GERALD DWORKIN, THE THEORY AND PRACTICE OF AUTONOMY 6 (1988). It is possible to distinguish, for example, between Razian, Kantian, hierarchical, reflective endorsement, historical, coherentist, and relational accounts of autonomy. See also COLBURN, supra note 104, at chapter 1 (giving an overview of different theories of autonomy).

[106] See Richard E. Ashcroft, Law and the Perils of Philosophical Grafts, 44 J. MED. ETHICS 72 (2018).

[107] See, e.g., Alberto Alemanno & Anne-Lise Sibony, Epilogue: The Legitimacy and Practicability of EU Behavioural Policy-Making, in NUDGE AND THE LAW 325, 326–33 (Anne-Lise Sibony & Alberto Alemanno eds., 2015); Fabrizio Esposito, Conceptual Foundations for a European Consumer Law and Behavioural Sciences Scholarship, in RESEARCH METHODS IN CONSUMER LAW: A HANDBOOK 1, 45 (Hans-Wolfgang Micklitz et al. eds., 2018); Fassiaux, supra note 3, at 1, 2–6; MARIJN SAX, BETWEEN EMPOWERMENT AND MANIPULATION 130–31 (2021); Daniel Susser et al., Online Manipulation: Hidden Influences in a Digital World, 4 GEO. L. TECH. REV. 1, 34–44 (2019); Karen Yeung, Nudge as Fudge, 75 MOD. L. REV. 122, 135 (2012).

[108] Whereas personal autonomy concerns an individual's capacity to live one's life according to one's own reasons and motives, moral autonomy concerns an individual's capacity to subject oneself to (objective) moral principles and to live one's life according to the right reasons and motives. See John Christman & Joel Anderson, Introduction, in AUTONOMY AND THE CHALLENGES TO LIBERALISM 1, 2 (John Christman & Joel Anderson eds., 2005). EU consumer law does not mandate that consumers adhere to (Kantian) moral principles. Hence, moral autonomy is not the appropriate philosophical framework for analysing EU consumer law.

[109] See Ashcroft, supra note 106, at 72.

[110] Christman, supra note 61, at 5.

[111] See DWORKIN, supra note 105, at chapter 1. See also Richard J. Arneson, Autonomy and Preference Formation, in IN HARM'S WAY: ESSAYS IN HONOR OF JOEL FEINBERG 42, 54 (Jules L. Coleman & Allan Buchanan eds., 1994) (distinguishing between the "Real Self condition" and the "Independence condition"); Christman & Anderson, supra note 108, at 3.

[112] See Lawrence Haworth, Autonomy and Utility, 95 ETHICS 5, 8 (1984).

## B. *THE INFORMATION PARADIGM AND ITS UNDERPINNING CONCEPTION OF AUTONOMY*

This Section advances the following two claims. First, it derives the definition of a model self-determined decision within the framework of EU consumer law from the information paradigm. Second, it posits that the information paradigm facilitates and the law regulating distorting external influences, such as provisions targeting dark patterns protects consumers' *ability* to make a model autonomous decision. These measures do not facilitate or protect the model self-determined decision itself. The Section explains the significance of this distinction.

The starting point for deriving a conception of autonomy from the information paradigm is the claim that providing consumers with information enables consumers to make an informed decision.[116] Establishing the link between mandated disclosure and an informed decision requires consumers' ability to make an informed choice based on this information. The information paradigm in EU consumer law assumes that this ability exists, a notion that has come under scholarly criticism.[117] Specifically, the information paradigm assumes that consumers are able to (i) pay attention to, read, and retain for long enough the information provided to them based on disclosure mandates; (ii) correctly understand this information; (iii) carefully consider this information as part of their decision-making process; and (iv) make their decisions based on this information rather than on other factors.[118]

---

[113] *See* DWORKIN, *supra* note 105, at 18. *See also* RAZ, *supra* note 61, at 372–73 (noting that autonomy as independence is associated with freedom from coercion and manipulation).
[114] *See, e.g.*, JOHN CHRISTMAN, THE POLITICS OF PERSONS 165–66 (2009); *see also* MARINA OSHANA, PERSONAL AUTONOMY IN SOCIETY 49–50 (2016).
[115] *See* ANDREW SNEDDON, AUTONOMY 87 (2013).
[116] *See, e.g.*, PRIIPs Regulation, *supra* note 92, art. 1 ("This Regulation lays down uniform rules on the format and content of the key information document . . . in order to enable retail investors to understand and compare the key features and risks of the PRIIP."); CCD II, *supra* note 87, art. 3(13) ("'pre-contractual information' means the information . . . which the consumer needs in order to be able to compare different credit offers and take an informed decision").
[117] Empirical evidence shows that decision-making in real life is often constrained by limited attention, limited time, and limited cognitive and decisional skills of consumers (*see, e.g.*, Gregory Crawford et al., *Consumer Protection for Online Markets and Large Digital Platforms*, 40 YALE J. ON REG. 101, 109–10 (2023); George Loewenstein et al., *Disclosure: Psychology Changes Everything*, 6 ANN. REV. ECON. 391–419 (2014)). This evidence limits the effectiveness of the information paradigm to empower consumers with mandated disclosure.
[118] Being able to pay attention to, read, retain, understand, consider, and make decisions based on information presupposes some degree of cognitive and decisional skills of consumers. This is expressed and specified in the competency condition of autonomy, which is an element of the self-determination condition of autonomy. *See, e.g.*, Christman & Anderson, *supra* note 108, at 3; *see also* SNEDDON, *supra* note 115, at 25–26.

Considering information before making a decision implies that (i) the information provided to consumers generates reasons in the process of decision-making for acting in one way rather than another and (ii) this process is a conscious process carried out with awareness. Consumers who consider and act upon information are aware of the reasons that drive their choices. They engage in a process of reflection (deliberation). The PRIIPs Regulation calls this process "understanding and use of information."[119] In short, the information paradigm assumes that consumers are able to make reflective decisions based on mandated information. This ability connects the information paradigm with autonomy theory and in particular the self-determination condition of autonomy. Even though different meanings of self-determination exist in philosophical discourse, one prominent theory of autonomy links self-determination to reflective decision-making on the basis of reasons.[120] Drawing upon the assumptions underlying the information paradigm, it is possible to specify a model self-determined decision for the purposes of EU consumer law as follows: A self-determined consumer decision is a decision based on mandated information, which the consumer correctly understands and which generates reasons in the decision-making process. Consumers are aware of these reasons that can explain and justify their decisions.

Even though the meaning of a model self-determined consumer decision can be derived from the information paradigm, it is important to distinguish between consumers' ability to make a model self-determined decision and the model decision itself. Recall that providing consumers with information *enables* consumers to make an informed decision. Likewise, the law regulating distorting external influences on consumer decision-making, such as provisions regulating dark patterns, protects consumers' *ability* to make an informed decision.[121] Translated into autonomy theory, the information paradigm enables consumers to make a model self-determined decision, and the provisions governing dark patterns protect consumers' ability to make a model self-determined decision. Neither the information paradigm nor the provisions governing dark patterns enforce, guarantee, or ensure that consumers actually make informed decisions or model autonomous decisions.[122] This is compatible with autonomy theory, which distinguishes between the capacity to act autonomously (autonomous persons) and exercising this capacity, that is acting autonomously (autonomous actions).[123] Being an autonomous person, that is having the ability to make a model self-determined decision, is compatible with individuals not exercising this ability in specific choice contexts and making a considerable number of decisions in a non-ideal way.[124]

---

[119] PRIIPs Regulation, *supra* note 92, recital 17.

[120] *See, e.g.*, Keith Lehrer, *Reason and Autonomy*, 20 SOC. PHIL. & POL'Y 177–98 (2003); GEORGE SHER, BEYOND NEUTRALITY: PERFECTIONISM AND POLITICS 48 (1997) ("exercising their will *on the basis of good reasons*" (emphasis added)).

[121] *See, e.g.*, DSA, *supra* note 5, art. 25(1) ("the ability of the recipients of their service to make free and informed decisions"); CRD, *supra* note 63, art. 16e(1) ("their [consumers who are recipients of their service] ability to make free and informed decisions"); *see also* UCPD, *supra* note 41, art. 2(e) ("'to materially distort the economic behaviour of consumers' means using a commercial practice to appreciably impair the consumer's ability to make an informed decision . . .").

[122] *Cf.* Luke Herrine, *What is Consumer Protection For?*, 34 LOY. CONSUMER L. REV. 240, 265 (2022) ("[P]roviding more information, better information, or simpler information does not necessarily guide consumers toward better choices.").

[123] *See* Beauchamp, *supra* note 104, at 61–62; DWORKIN, *supra* note 105, at 19–20; FADEN & BEAUCHAMP, *supra* note 91, at 235–37.

EU consumer law protects consumers as autonomous persons rather than the model autonomous decision itself. Both the information paradigm and the provisions pertaining to dark patterns recognize consumers' freedom to ignore mandated information and to base their decision on factors like advertising, online reviews, emotions, recommendations from friends or influencers, and so on.[125] Even though the information paradigm privileges reflection as a decision-making process over other, non-reflective processes, it does not exclude other possible motivations of consumer decisions. To this extent, the information paradigm accepts consumers' *own* (conscious or subconscious) principles of decision-making and consumers' own decision to decide unreflectively in specific choice contexts as an act of an autonomous person. This perspective aligns with the view in autonomy scholarship that the thought processes that constitute the self involve conscious and subconscious processes.[126] Support for this perspective is also provided by Double, who has argued that individuals are autonomous if they choose according to their own subjectively preferred "individual management style," which can include non-reflective decision-making.[127] Similarly, Alemanno and Sibony have argued that the extent to which a consumer decides reflectively or unreflectively is a decision about how consumers manage their scarce deliberative resources ("mental bandwidth"), a decision they perceive as integral to one's identity.[128] The information paradigm acknowledges the diverse approaches individuals take in allocating their "mental bandwidth," thus respecting consumers as autonomous persons. Embedded within the information paradigm in EU consumer law is the acknowledgement that providing consumers with information may not always lead to informed decisions. This acknowledgement in and of itself is neither a flaw of the information paradigm nor does it diminish its effectiveness when measured against the benchmark of consumer autonomy. Instead, it protects consumers as autonomous persons.[129]

What needs to be distinguished from perceiving a deviation from a model self-determined decision as an act of an autonomous person is a deviation from a model self-determined decision caused by an external influence. In the latter case, deciding unreflectively may not be an act of an autonomous person about how to manage their scarce deliberative resources. A critical distinction exists between non-reflective decision-making based on one's own decision-making principles, one's own individual management style, and non-reflective decision-making induced by a distorting external influence. For instance, external influences which cause consumers to ignore mandated information may violate consumers' ability to make a model self-determined decision. Such influences may violate the procedural independence condition of autonomous choice and may, therefore, warrant regulation.[130]

---

[124] *See* JONATHAN PUGH, AUTONOMY, RATIONALITY, AND CONTEMPORARY BIOETHICS 200–01 (2020). *Cf.* DWORKIN, *supra* note 105, at 17; Viktor Ivanković & Bart Engelen, *Market Nudges and Autonomy,* ECON. & PHIL. 9 (2023), https://doi.org/10.1017/S0266267122000347 [https://perma.cc/RFV8-EYX9] ("autonomy is not necessarily undermined if agents decide against or neglect to pay attention to their preference formation processes").

[125] *Cf.* Jules Stuyck et al., *Confidence Through Fairness? The New Directive on Unfair Business-to-Consumer Commercial Practices in the Internal Market*, 43 COMMON MKT. L. REV. 107, 125 (2006) (for the UCPD, *supra* note 41).

[126] *See* SNEDDON, *supra* note 115, at 81; *see also* CHRISTMAN, *supra* note 114, at 140.

[127] *See* Richard Double, *Two Types of Autonomy Accounts*, 22 CAN. J. PHIL. 65, 68–73 (1992).

[128] *See* Alemanno & Sibony, *supra* note 107, at 330–32.

[129] This point is not undermined by empirical evidence showing that consumer decision-making in real life is often constrained by limited attention, limited time, and limited cognitive and decisional skills (*see* the references in *supra* note 117). This empirical evidence suggest that consumers may lack the ability to make a model self-determined decision. The absence of the ability to make a model self-determined decision must be distinguished from possessing this ability but not exercising it in certain contexts.

### C. *THE EVOLUTION OF THE INFORMATION PARADIGM AND ITS UNDERPINNING CONCEPTION OF AUTONOMY*

After specifying the meaning of autonomous consumer decision-making based on the EU's information paradigm in the preceding Section, this Section contends that the information paradigm, along with its underpinning conception of autonomy, have started to evolve by taking into account empirical insights about consumer decision-making. Empirical evidence has started to shape the meaning of consumer autonomy by modifying two assumptions underlying the information paradigm. This Section also explains how this development affects the regulation of dark patterns.

One assumption underlying the information paradigm is that a consumer needs to have *all* relevant information before making a decision.[131] Information provided to consumers generates reasons in their decision-making process. Hence, this assumption is supported by the view in autonomy theory that "the more reasons . . . that one is capable of seeing and understanding, the more fully one can claim one's choice to be one's own."[132] Helleringer and Sibony have argued that this assumption has contributed to the expansion of information obligations based on the "credo . . . that more information is always better for consumers."[133] While it is evident that certain recent EU legislation still adheres to this credo,[134] other legislative acts suggest that the information paradigm is—gradually, and certainly not uniformly—moving away from this credo. One example is key information documents in modern financial consumer law like the PRIIPs KID. These documents illustrate that a model autonomous consumer decision does not have to be based on full or all relevant information. It is sufficient if the consumer considers the essential information. This resonates with accounts of autonomy that require substantial but not full understanding for autonomous decision-making.[135] The PRIIPs Regulation acknowledges that "unless the information is short and concise there is a risk that [retail investors] will not use it."[136]

This evolution of the information paradigm recognizes the issue of information overload in consumer markets. Information overload can have a detrimental effect on an individual's ability to make autonomous decisions as it may reduce an individual's understanding of the available options.[137] Empirical research demonstrates that excessive information can reduce the ability to critically reflect and the quality of decision-making.[138] This is not only significant for the information paradigm itself but also for the provisions regulating dark patterns. That is because creating information overload in online choice architectures may violate consumers' ability to make a model autonomous decision.[139]

---

[130] *See, in detail, infra* Part III.

[131] *See, e.g.*, Geneviève Helleringer & Anne-Lise Sibony, *European Consumer Protection Through the Behavioral Lens*, 23 COLUM. J. EUR. L. 607, 617–18 (2017); MALEK RADEIDEH, FAIR TRADING IN EC LAW 198, 207 (2005); Annette Nordhausen Scholes, *Behavioural Economics and the Autonomous Consumer*, 14 CAMBRIDGE Y.B. EUR. LEGAL STUD. 297, 306 (2012).

[132] SUSAN WOLF, FREEDOM WITHIN REASON 144 (1993).

[133] Helleringer & Sibony, *supra* note 131, at 622.

[134] *See* Arno R. Lodder & Jorge Morais Carvalho, *Online Platforms: Towards an Information Tsunami with New Requirements on Moderation, Ranking, and Traceability*, 33 EUR. BUS. L. REV. 537–56 (2022) (discussing the Omnibus Directive (EU) 2019/2161).

[135] *See, e.g.*, FADEN & BEAUCHAMP, *supra* note 91, at chapter 9.

[136] PRIIPs Regulation, *supra* note 92, recital 15.

A second assumption underlying the information paradigm relates to the presentation of information. The traditional information paradigm assumes that a consumer is able to correctly understand and use the information once it is provided to the consumer, which is why it is not necessary to prescribe its specific presentation in detail. Hence, presentation requirements used to focus on language and remain at the level of platitudes like "clear and comprehensible."[140] A sign that this assumption is evolving is consumer laws which regulate the presentation of mandated information in detail. These laws are informed by empirical evidence showing that the presentation of information can significantly affect how consumers understand and use information.[141] Admittedly, this regulatory evolution remains tentative and is most developed in sector-specific EU legislation like food legislation or financial regulation.[142] A case in point is the detailed presentation requirements for the PRIIPs KID.[143] These requirements are not limited to the use of language but also include the context in which information has to be presented. For example, the PRIIPs Regulation uses salience effects when mandating that: "The title 'Key Information Document' shall appear prominently at the top of the first page of the key information document."[144] Salience effects typically work at the non-reflective level.[145] For example, consumers do not usually reflect on whether or to what extent the position, size, or color of an agree or buy button influences their decision-making. Consumers do not usually reflect on the extent to which the order of items in search results affects their clicking behavior. Consumers do not usually reflect on the extent to which the vivid presentation of irrelevant information affects their decision to purchase a service or a product.

---

[137] See PUGH, supra note 124, at 145, 169.

[138] See, e.g., Elena Reutskaja et al., Cognitive and Affective Consequences of Information and Choice Overload, in ROUTLEDGE HANDBOOK OF BOUNDED RATIONALITY 625, 628–33 (Riccardo Viale ed., 2020).

[139] See infra Part III.B.1.

[140] See, e.g., CRD, supra note 63, arts. 5(1), 6(1) and 6a(1); Directive 2000/31/EC of the European Parliament and of the Council of 8 June 2000 on Certain Legal Aspects of Information Society Services, in Particular Electronic Commerce, in the Internal Market (Directive on Electronic Commerce), art. 10(1), 2000 O.J. (L 178) 1. Cf. Joasia Luzak et al., ABC of Online Consumer Disclosure Duties: Improving Transparency and Legal Certainty in Europe, 46 J. CONSUMER POL'Y 307, 310–12 (2023) (criticizing broadly and vaguely set information presentation requirements for their uncertainty).

[141] See, e.g., Luzak et al., supra note 140, at 326 (arguing that moving from textually to visually enhanced information presentation dramatically improves consumer understanding); Jacob L. Orquin et al., Visual Biases in Decision Making, 40 APPLIED ECON. PERSP. & POL'Y 523–37 (2018); Ognyan Seizov et al., The Transparent Trap: A Multidisciplinary Perspective on the Design of Transparent Online Disclosures in the EU, 42 J. CONSUMER POL'Y 149, 159–164 (2019) (providing a concise overview of the empirical research).

[142] See, e.g., Regulation (EU) No. 1169/2011, arts. 12–13, 2011 O.J. (L 304) 18; Directive 2014/17/EU (Mortgage Credit Directive), annex II, 2014 O.J. (L 60) 34; Commission Delegated Regulation (EU) 2017/565, art. 44, 2017 O.J. (L 87) 1. Cf. Jablonowska & Tagiuri, supra note 13, at 20 (noting the expansion of explicit provisions on disclosure modalities, reflecting the growing attention that EU policymakers devote to information presentation requirements that facilitate understanding).

[143] See, e.g., PRIIPs Regulation, supra note 92, art. 8.

[144] Id. at art. 8(1).

[145] See infra Part III.A. See also KEITH STANOVICH, RATIONALITY AND THE REFLECTIVE MIND 104, 111 (2010).

The PRIIPs Regulation recognizes that factors that consumers do not necessarily reflect on, such as how salient mandated information is presented, can help consumers use and understand mandated information. Recognizing that the manner in which information is presented can significantly affect consumers' understanding and use of information also suggests that a wider range of presentation formats could potentially distort consumers' understanding and use of information. This is one factor that explains why the regulation of distortive private influences on autonomous consumer decision-making such as dark patterns has intensified in modern consumer law, particularly in online choice environments where businesses constantly experiment with design choices in order to maximize their profits.

To conclude this Section, I have argued that the EU's information paradigm and its underpinning conception of autonomy are evolving based on empirical evidence about consumer decision-making. This has been a slow process so far, and behavioral evidence about consumer biases, consumer heuristics, and the effects of the presentation of information on consumer decision-making has certainly not led to an overhaul of the normative conception of autonomy which is based on mandated information, reflection, and awareness.

## D. *AUTONOMOUS AND RATIONAL CONSUMER DECISIONS*

So far, I have argued that the information paradigm is underpinned by a specific conception of autonomy. This Section addresses the relationship between autonomous and rational consumer decision-making in this conception of autonomy. The Section challenges the dominant position in the literature which holds that the information paradigm assumes that consumers are rational economic actors.[146] This assumption creates a link between the information paradigm and the neoclassical model of rationality (rational choice theory). Whereas rational choice theory concerns choice outcomes and how they relate to each other but is agnostic with respect to how choices are being made,[147] the model self-determined decision that underlies the information paradigm is not concerned with choice outcomes but with the process of understanding and reasoning leading up to a consumer decision. This distinction between decision outcome and decision process is also contained in the law regulating distorting external influences on consumer decision-making, such as the provisions regulating dark patterns or the UCPD. This law does not protect the informed decision itself but consumers' *ability* to make an informed decision. The consumer is free to ignore all information and make an unwise decision that is detrimental to the consumer's welfare, provided that the consumer's decision-making process was not appreciably impaired.[148]

---

[146] *See* Incardona & Poncibò, *supra* note 46; Mak, *supra* note 46; Oehler & Wendt, *supra* note 102; Stănescu, *supra* note 46.

[147] *See* B. Douglas Bernheim, *The Psychology of Judgment and Decision Making: What's in it for Economists?*, in NEUROECONOMICS: DECISION MAKING AND THE BRAIN 115, 119 (Paul W. Glimcher et al. eds., 2008).

[148] *See* Stuyck et al., *supra* note 125, at 125 (for the UCPD, *supra* note 41).

Furthermore, the PRIIPs Regulation requires the European Commission to review the Regulation.[149] This review shall assess whether "the measures introduced have improved the average retail investor understanding of PRIIPs and the comparability of the PRIIPs."[150] This suggests that a better understanding is intrinsically valuable, irrespective of the welfare-related consequences of this understanding.

I am not denying that there is a link between autonomous and rational decision-making. In fact, many philosophers contend that a rationality requirement of some sort is a necessary condition for autonomous decision-making.[151] The information paradigm assumes that consumers are able to decide on the basis of reasons, and one view in philosophy understands rationality as responding adequately to reasons.[152] According to this view, rationality is an attribute of the decision-making process rather than an attribute of the choice outcome. A consumer who decides on the basis of reasons responds to reasons. Hence, a rationality requirement is incorporated in the model self-determined consumer decision. Since this requirement links to process—rather than outcome—rationality, autonomous decision-making cannot be equated with an economically rational choice. It follows that the information paradigm does not assume that consumers are rational economic actors.

While the model self-determined decision includes a rationality requirement, the concept of autonomy in EU consumer law cannot be equated with rational self-governance driven solely by reasons, despite the existence of such theories of ideal autonomy in philosophical discourse.[153] While rationality requires a decision-making process that responds to reasons (as opposed to other motivations, for example), the conception of autonomy underpinning the information paradigm relates to consumers' ability to respond to reasons (to decide on the basis of reasons). A consumer who decides non-reflectively and ignores mandated information does not exercise the ability to respond to reasons. Nevertheless, the consumer's decision-making in such a case is regarded as decision-making of an autonomous person if the consumer had the ability to consider the information and decide on the basis of reasons. This divergence between the conception of autonomy in EU consumer law and process rationality does not pose a conceptual dilemma. This is because the rationality requirement in the autonomy benchmark does not have to be identical with and can deviate from one's understanding of process rationality.[154] Lindley has explained this distinction between rationality and autonomy as follows: "[W]hilst autonomy is primarily a matter of authorship, rationality is essentially a matter of acceptability."[155]

---

[149] *See* PRIIPs Regulation, *supra* note 92, art. 33.

[150] *Id.* at recital 36.

[151] *See* Joel Anderson, *Autonomy*, in INTERNATIONAL ENCYCLOPEDIA OF ETHICS 442–58 (Hugh LaFollette ed., 2013); John Christman, *Constructing the Inner Citadel: Recent Work on the Concept of Autonomy*, 99 ETHICS 109, 115–16 (1988); RICHARD LINDLEY, AUTONOMY 13–70 (1986) (all three discussing different perspectives within the literature regarding the necessity and nature of a rationality requirement for autonomous decision-making).

[152] *See, e.g.*, Nora Heinzelmann, *Rationality is Not Coherence*, 74 PHIL. Q. 312–32 (2024); BENJAMIN KIESEWETTER, THE NORMATIVITY OF RATIONALITY ch.7 (2017).

[153] *See, e.g.*, SHER, *supra* note120, at 48–49 (1997). *Cf.* COLBURN, *supra* note 104, at 7 (discussing these theories).

[154] *See* Arneson, *supra* note 111, at 47. *Cf.* Christopher Mills, *Manipulation and Autonomy*, in THE ROUTLEDGE HANDBOOK OF AUTONOMY 223, 227 (Ben Colburn ed., 2022) ("Autonomy cannot plausibly require agents to possess perfect information and only act according to the best reasons.").

[155] LINDLEY, *supra* note 151, at 21.

E.  CONCLUSION

This Part has specified the meaning of autonomy in EU consumer law. It has developed a normative conception of autonomous decision-making based on the information paradigm. The information paradigm enables consumers to make model self-determined decisions. It does not assume that consumers are rational economic actors. A model self-determined decision is a decision based on mandated information, which the consumer correctly understands and which generates reasons in the decision-making process. Consumers are aware of these reasons that can explain and justify their decision. The provisions regulating dark patterns in EU consumer law protect consumers' ability to make model self-determined decisions. They protect the procedural independence of consumer decision-making.

III.  DARK PATTERNS AS VIOLATIONS OF AUTONOMY

This Part explicates when dark patterns in online choice architectures violate consumer autonomy. The first Section clarifies the contentious relationship between consumer biases, consumers' non-reflective decision-making, and autonomy violations. The second Section develops six categories of autonomy violations in b2c relationships which are particularly relevant for assessing and regulating dark patterns in online choice architectures in EU consumer law. This Section also shows how these categories apply to specific dark pattern practices that are commonly discussed in the literature and by policymakers such as drip pricing, subscription traps, default settings that maximize the collection of data, or website designs that effectively hide fees.

A.  CONSUMER BIASES, NON-REFLECTIVE DECISION-MAKING, AND AUTONOMY VIOLATIONS

This Section contends that not all online choice architectures that trigger[156] a consumer bias violate consumer autonomy. To develop this claim, it is fitting to start with the dominant psychological framework for explanations of cognitive biases: Dual-process theory. This theory classifies thought processes in a dual-process model, System 1 and System 2. Whereas System 2 is characterized by reflective reasoning, System 1 is characterized by non-reflective, automatic, and intuitive forms of processing.[157] Heuristic processes are often equated with System 1 decision-making.[158] In earlier writings on the psychology of reasoning, cognitive biases were clearly associated with non-reflective decision processes and were said to result from people's use of heuristics.[159] If biases are caused by consumers' use of non-reflective heuristics and if an online choice architecture triggers an observable consumer bias, it can be assumed that the triggering of the bias was achieved by influencing consumers' non-reflective decision-making processes which the consumer is not aware of and which, therefore, do not constitute reasons that can explain and justify the consumer's decision. This assumption appears to be the backbone of the literature arguing that autonomy is undermined when biases and heuristics are present in human decision-making.[160]

---

[156] In this Section, "triggering" a consumer bias encompasses the exacerbation of consumer biases.
[157] *See* Jonathan St. B. T. Evans, *Bounded Rationality, Reasoning and Dual Processing, in* ROUTLEDGE HANDBOOK OF BOUNDED RATIONALITY 185, 188 (Riccardo Viale ed., 2020); DANIEL KAHNEMAN, THINKING, FAST AND SLOW 20 (2012).
[158] *See, e.g.*, STANOVICH, *supra* note 145, at 29–31.
[159] *See* Jonathan St. B. T. Evans, *Dual-Process Theories of Deductive Reasoning: Facts and Fallacies, in* THE OXFORD HANDBOOK OF THINKING AND REASONING 115, 126–27 (Keith J. Holyoak & Robert G. Morrison eds., 2012).

More recent research on the psychology of reasoning shows, however, that it is a fallacy to assume that only non-reflective System 1 processes are responsible for cognitive biases. Evans and Stanovich have argued that reflective System 2 processes can also be responsible for cognitive biases in some circumstances.[161] Nonetheless, there is no reason to throw the baby out with the bathwater. Stanovich, for example, maintains that "most social and cognitive biases operate unconsciously."[162] He argues that well-known biases like salience, affect, framing, and anchoring bias are based on heuristic, non-reflective processing.[163]

When an online choice architecture triggers a consumer bias, we may thus continue to assume that this triggering was achieved by influencing consumers' non-reflective decision-making processes. It does not however follow that every successful external influence on consumers' non-reflective decision-making processes amounts to a violation of the procedural independence condition of autonomy and violates consumers' ability to make a model self-determined decision.[164] Not every external influence on consumer decision-making must appeal to reason and reflection. While this standpoint is not uncontroversial,[165] an opposing perspective could deem a wide array of presentation and advertising techniques in online choice architectures as ethically wrong. Much of modern advertising appeals to non-reflective decision-making like consumers' emotions. Advertisements may not appeal to reflection at all, and this is socially accepted since external appeals to non-reflective decision-making are a pervasive (relational) feature of human life.[166] Even though this assessment is subject to debate,[167] it is contained in the UCPD, which recognizes that the "advertising practice of making exaggerated statements or statements which are not meant to be taken literally" is "common and legitimate."[168] This normative assessment also applies in the context of the Digital Services Act.[169]

---

[160] See, e.g., Jennifer S. Blumenthal-Barby, Biases and Heuristics in Decision Making and Their Impact on Autonomy, 16 AM. J. BIOETHICS 5–15 (2016); Abraham P. Schwab, Formal and Effective Autonomy in Healthcare, 32 J. MED. ETHICS 575–79 (2006).

[161] See Evans, supra note 159, at 126–27; Jonathan St. B. T. Evans & Keith E. Stanovich, Dual-Process Theories of Higher Cognition: Advancing the Debate, 8 PERSP. ON PSYCHOL. SCI. 223, 229 (2013).

[162] STANOVICH, supra note 145, at 112.

[163] See id. at 104–06, 111.

[164] See, e.g., Jennifer S. Blumenthal-Barby, A framework for Assessing the Moral Status of "Manipulation", in MANIPULATION: THEORY AND PRACTICE 121, 126–27 (Christian Coons & Michael Weber eds., 2014); Philipp Hacker, Nudging and Autonomy: A Philosophical and Legal Appraisal, in RESEARCH METHODS IN CONSUMER LAW: A HANDBOOK 77, 96 (Hans-Wolfgang Micklitz et al. eds., 2018).

[165] See, e.g., Douglas MacKay & Alexandra Robinson, The Ethics of Organ Donor Registration Policies: Nudges and Respect for Autonomy, 16 AM. J. BIOETHICS 3, 7 (2016) ("The use of reason-bypassing nonargumentative influence is disrespectful of people's autonomy since it violates (2), corrupting people's decision-making processes by working around or bypassing their deliberative capacities . . .").

[166] Cf. Jan Christoph Bublitz & Reinhard Merkel, Autonomy and Authenticity of Enhanced Personality Traits, 23 BIOETHICS 360, 368 (2009) (arguing that communication in daily life always involves non-conscious factors and that sales communication often contains socially accepted subconscious interventions).

[167] See, e.g., Roger Crisp, Persuasive Advertising, Autonomy, and the Creating of Desire, 4 J. BUS. ETHICS 413–18 (1987); Richard L. Lippke, Advertising and the Social Conditions of Autonomy, 8 BUS. & PROF. ETHICS J. 35–58 (1989).

Furthermore, any choice architecture will have some influence on non-reflective decision-making processes. That is because human decision-making always involves non-reflective factors,[170] which also means that it is impossible (and undesirable) to completely eliminate non-reflective factors from human decision-making. Similarly, Cohen has argued that even "rational persuasion does not happen in a vacuum—it must take place within some dialogical setting, and the contingencies of that setting would count as nonrational influences."[171] The challenge resides in clearly delineating the boundary separating external influences on consumers' non-reflective decision-making that violate autonomy from those that do not. The next Section elaborates on refining this demarcation.

## B.  SIX CATEGORIES OF AUTONOMY VIOLATIONS

This Section develops six categories of autonomy violations in b2c relationships. These categories have two purposes. First, they uncover and make explicit the autonomy violation which is targeted by existing provisions regulating dark patterns in EU consumer law. This helps interpreting these provisions, applying them in borderline cases, and making them more effective tools of consumer protection. In particular, the six categories of autonomy violations can be used to specify the vague criteria employed by the provisions governing dark patterns in the DSA, the CRD, and the Data Act.[172] The European Commission should also rely on these categories when issuing guidelines on how the prohibition of dark patterns in the DSA applies to the listed practices in DSA art. 25(3).

The second purpose of the categories of autonomy violations is to abstract from specific regulated dark pattern practices. Currently unregulated dark pattern practices falling under these categories of autonomy violations have significant similarity to already regulated practices. Consequently, these unregulated dark patterns are not only ethically wrong due to their autonomy violation, but they could also warrant regulatory intervention. I do not claim here that all instances of dark patterns in online choice architectures which are captured by these categories should be regulated. Such a claim is not possible since determining whether the state can intervene in dark patterns requires a balancing exercise weighing up conflicting normative arguments.[173] The violation of personal autonomy does not change this assessment since autonomy is a relative and not an absolute value. What I am claiming is that these categories should be used by EU policymakers when deciding whether to regulate other cases of dark patterns.

---

[168] UCPD, *supra* note 41, art. 5(3) sentence 2.

[169] *See* DSA, *supra* note 5, recital 67 ("Legitimate practices, for example in advertising, that are in compliance with Union law should not in themselves be regarded as constituting dark patterns.").

[170] *See* EVANS, *supra* note 36, at 167; JONATHAN ST. B. T. EVANS, THINKING TWICE: TWO MINDS IN ONE BRAIN 2–4, 107, 187 (2010); Tanya L. Chartrand & Gavan J. Fitzsimons, *Nonconscious Consumer Psychology*, 21 J. CONSUMER PSYCHOL. 1 (2011).

[171] Shlomo Cohen, *A Philosophical Misunderstanding at the Basis of Opposition to Nudging*, 15 AM. J. BIOETHICS 39, 40 (2015).

[172] *See* DSA, *supra* note 5, art. 25(1) ("distorts or impairs the ability of the recipients of their service to make free and informed decisions."), CRD, *supra* note 63, art. 16e(1) ("distorts or impairs their [consumers who are recipients of their service] ability to make free and informed decisions"), Data Act, *supra* note 71, art. 4(41c) ("subverting or impairing the autonomy, decision-making or choices of the user"), and Data Act, *id.* at art. 6(2)(a) ("subverting or impairing the autonomy, decision-making or choices of the user").

The analysis in this Section is inductive and legalist. It is based on the existing provisions regulating dark patterns in EU consumer law, which were analyzed in Part I of this Article. Alternatively, categories of autonomy violations could also be developed deductively based on a theory which specifies the line between autonomy-violating and non-violating external influences on consumers decision-making. EU consumer law does not contain such a theory, and using a philosophical theory as the bedrock of the analysis may lead to categories of autonomy violations that are incompatible with existing EU consumer law.

### 1.  *Category 1: Undermining of Mandated Information*

*Dark patterns in online choice architectures violate consumers' ability to make a model self-determined decision if they influence consumers' decision-making in such a way that consumers ignore or misunderstand mandated information.*

This effect of the choice architecture on consumer decision-making actively undermines legislative efforts (the provision and presentation of mandated information), which aim to enable consumers to use and understand the information that is meant to form the basis of consumers' decision-making. The PRIIPs KID, for example, protects consumers from this category of autonomy violations by taking control of the presentation of mandated information. Category 1 also captures the following "design elements"[174] in online choice architectures if these design elements cause consumers to misunderstand or ignore mandated information: (i) the creation of information overload (for example, burying key terms of service within dense terms & conditions documents for customers); (ii) the salient presentation of irrelevant information; and (iii) confusing language like double negatives. Category 1 is particularly relevant for assessing online choice architectures that attempt to hide the existence of fees, for example button designs that consumers are unlikely to click on, the burying of information about fees between blocks of salient information, or displaying fees only at the end of a long online document which requires considerable scrolling down.[175]

Category 1 can also apply to certain cases of drip pricing, which is prevalent in live events, hotel, and airline industries. Drip pricing refers to the practice of gradually revealing additional, optional, or mandatory fees during the checkout process as opposed to disclosing such fees up front, making the total price of a product or service higher than the low headline price which was revealed to the consumer at the beginning of the purchase process. For example, a hotel website might advertise a room rate of $100 per night, but additional mandatory fees for resort amenities, cleaning, or Wi-Fi are added on top of the headline price and made transparent just before finishing the booking process. Even though it is often said that drip pricing takes advantage of cognitive biases and heuristics,[176] multiple behavioral theories can explain the effects of drip pricing on consumer decision-making, and there is currently no consensus as to which explanation is superior or best.[177] The dominant theory is anchoring and adjustment. Biased consumers anchor on the headline price and later in the purchasing process adjust insufficiently for the additional price increments and, hence, underestimate the total price.[178] If the total price is only revealed to the consumer at the end of the booking process, consumers may also feel invested in the process and complete the purchase to not waste their time and effort (sunk cost fallacy).[179]

---

[173] *Cf.* Brenncke, *supra* note 99, at 11 (for state intervention in commercial practices exploiting consumer behavioral biases). For example, the prohibition of dark patterns on online interfaces of online platforms according to DSA, *supra* note 5, art. 25(1) does not apply to providers of online platforms that qualify as micro or small enterprises (*see* DSA, *supra* note 5, art. 19(1)).

[174] Design elements in online choice architectures cover the structure, design, function, or manner of operation of the online choice architecture. *See* DSA, *supra* note 5, art. 25(1); CRD, *supra* note 63, art. 16e(1); Data Act, *supra* note 71, arts. 4(4) and 6(2)(a).

[175] Categories 2 and 3 of autonomy violations are also able to capture certain online designs attempting to hide the existence of fees. *See infra* Parts III.B.2 and III.B.3.

Studies on drip pricing typically focus on the economic costs associated with this practice and often overlook or provide minimal insights into the implications of drip pricing for autonomy.[180] This trend is also observed among regulators. For instance, in October 2023, the U.S. Federal Trade Commission proposed to prohibit commercial practices that misrepresent the total costs of goods or services by omitting mandatory fees from advertised prices, such as the dripping of mandatory fees.[181] While the Commission thoroughly discusses the economic rationale for the proposed prohibition, it does not elaborate on the impact of drip pricing on consumer autonomy. Why might drip pricing violate autonomy? Since the total price of a good or service is a piece of mandated information,[182] drip pricing is a Category 1 autonomy violation if its effect on consumer decision-making is that consumers do not correctly understand the total price of the product or service or ignore the total price (and use the headline price instead) when making purchasing decisions. Admittedly, Category 1 has a limited ability to capture drip pricing if the total price is shown prominently at the last stage of the booking process, but other categories of autonomy violations may still apply to drip pricing.

---

[176] Other complex pricing strategies have also been linked to the exploitation of consumer behavioural biases. *See* van Loo, *supra* note 13, at 220–27.

[177] *See* Anne-Lise Sibony, *Did You Say 'Theories of Choice'? On the Limited and Variable Appetite for Theories in Consumer Law*, *in* THEORIES OF CHOICE: THE SOCIAL SCIENCE AND THE LAW OF DECISION MAKING 115, 127–28 (Stefan Grundmann & Philipp Hacker eds., 2021). *See also* Shelle Santana et al., *Consumer Reactions to Drip Pricing*, 39 MARKETING SCI. 188, 206–08 (2020) (arguing that multiple behavioral factors in combination cause the effect of drip pricing on consumer behavior).

[178] *See* Gorkan Ahmetoglu et al., *Pricing Practices: A Critical Review of Their Effects on Consumer Perceptions and Behaviour*, 21 J. RETAILING & CONSUMER SERV. 696, 697–99 (2014).

[179] *See* Mathur et al., *supra* note 1, at 81:13.

[180] *See, e.g.*, Philipp Brunner & Christian Zihlmann, *Drip Pricing, After-sales, and Sequential Buying with Behavioral Consumers* (2023), https://christianzihlmann.github.io/files/drippricing.pdf [https://perma.cc/37K7-FC5Y]; Alexander Rasch et al., *Drip Pricing and its Regulation: Experimental Evidence*, 176 J. ECON. BEHAV. & ORG. 353–70 (2020); Santana et al., *supra* note 177. *See also* David Adam Friedman, *Regulating Drip Pricing*, 31 STAN. L. & POL'Y REV. 51–102 (2020) (only assessing the economic impact of drip pricing without discussing the implications of drip pricing for autonomy).

[181] Federal Trade Commission, Trade Regulation Rule on Unfair or Deceptive Fees, 88 Fed. Reg. 216, 77420 (Nov. 9, 2023) (to be codified at 16 C.F.R. pt. 464).

[182] *See* CRD, *supra* note 63, art. 6(1)(e).

### 2. *Category 2: Deception*

*Dark patterns in online choice architectures violate consumers' ability to make a model self-determined decision if they cause consumers to hold false beliefs that form the foundation of their decision-making.*

Category 2 of autonomy violations is derived from DSA art. 25(1), which prohibits providers of online platforms to use online interface designs that deceive consumers. Equivalent prohibitions exist in the CRD and the Data Act.[183] The wording of DSA art. 25(1) implies that deception is a material distortion or impairment of consumers' ability to make free and informed decisions. Hence, a deceptive interface design on online platforms is an autonomy violation that warrants legal intervention. To deceive someone is to cause them to hold false beliefs.[184] Category 2 has a significant overlap with Category 1 of autonomy violations if the false belief relates to mandated information.

Category 2 of autonomy violations becomes pertinent in cases involving biased consumers, where an online choice architecture is able to induce a false belief due to a consumer bias that the choice architecture has triggered or exacerbated. For example, Category 2 captures the use of confusing language like a double negative in a consent banner which steers consumers towards agreeing with the sharing of data with third parties even though consumers falsely believe that they have rejected it. Category 2 also applies to online interface designs that ask consumers to agree to a certain policy and make highly salient the option to agree compared to the option to disagree if this design causes biased consumers, who are susceptible to salience effects, to hold the false belief that there is no option to reject the policy.[185]

Furthermore, Category 2 captures drip pricing if the design of the online choice architecture induces the false belief in biased consumers, for example due to salience or framing effects, that the initial (headline) price is the total price of the product or service. This evaluation remains unaffected even if the false belief is corrected at a later point in the purchasing process. This correction does not remove the autonomy violation that occurs when the consumer decides to start the time-consuming purchasing process based on the belief that the headline price is the total price. The relevant consumer decision is the transactional decision to start and complete the purchasing process with the headline price rather than the later decision to conclude the purchasing process at a different total price.

### 3. *Category 3: Inducing Contractual Agreements Without Reflection*

*Dark patterns in online choice architectures violate consumers' ability to make a model self-determined decision if they cause consumers to enter into a specific contractual agreement with a business without reflection about its substance and content.*

Category 3 specifies the autonomy violation that is the focus of the regulation of default options in the CRD and the CCD II. Both legislative acts regulate the use of default options such as pre-ticked boxes in (online)

---

choice environments in order to prevent the exploitation of consumers' behavioral biases.[186] It has been suggested in the literature that default options like pre-ticked boxes typically work by bypassing a person's reasoning capacities and awareness.[187] Consumers who choose a default option passively and unreflectively simply because it is the default do not reflect on the substance and content of their decision. Consumer's ability to make a model self-determined decision is violated in this scenario if the choice architect set the default option and thereby caused consumers to deviate from a model self-determined decision.

Both CRD art. 22 and CCD II art. 15 also capture the use of pre-ticked boxes if such boxes are transparent and even if consumers reflect on the substance and content of the agreement. However, this does not imply that both provisions adopt a normative theory other than autonomy for regulating such cases. Instead, both provisions contain the, admittedly controversial,[188] legislative generalization that pre-ticked boxes in (online) choice environments trigger consumer biases and bypass reflective decision-making processes. In the light of this legislative generalization, the following example also falls under Category 3: the use of pre-ticked boxes in online choice environments that default consumers into the automatic renewal of a product or service, which incurs subscription costs to the consumer.

Category 3 of autonomy violations also captures the following four examples of design choices in online choice environments. The first example is the sneaking of products into consumers' online shopping carts without their knowledge. This practice takes advantage of consumers' default effect cognitive bias.[189] A second example is a free trial that is followed by monthly subscription charges if the free trial is not cancelled in time, which is only disclosed by the statement that "terms and conditions apply." Biased consumers who do not read the lengthy terms and conditions document are not aware of their agreement to a monthly subscription when agreeing to the free trial. A third example is a free trial that converts to a paid subscription only because there is a pre-ticked checkbox with that effect when consumers sign up to the free trial. A fourth example is the following scenario: A green button that normally advances a consumer to the next level in a computer game is suddenly replaced with a green button that initiates an in-game purchase based on the single click of the button. A biased consumer who clicks the green button out of habit in order to get to the next level of the computer game is not aware that the click initiates the purchase. This last example is also a Category 2 autonomy violation. In all of the examples discussed in this Section, the consumer incurs financial costs which the consumer is not aware of, and this financial harm adds normative weight in favor of regulating these autonomy violations.

---

[186] *See supra* Part I.F.
[187] *See* Jennifer S. Blumenthal-Barby, *Between Reason and Coercion: Ethically Permissible Influence in Health Care and Health Policy Contexts*, 22 Kennedy Inst. Ethics J. 349, 349 (2012); MacKay & Robinson, *supra* note 165, at 6; Smith et al., *supra* note 38.
[188] *See* Sunstein, *supra* note 38, at 17–24 (explaining that (i) not all defaults trigger consumer biases or unreflective decision-making and (ii) consumers may opt for the default option for several reasons).
[189] *See* Mathur et al., *supra* note 1, at 81:13.

4.  *Category 4: Negative Friction*

   *Dark patterns in online choice architectures violate consumers' ability to make a model self-determined decision if they create unreasonable time, decision effort, or emotional costs for pursuing or adhering to a particular decision.*

   Such a design of the online choice architecture creates friction in the decision-making process, and it is this friction which effectively deters consumers from pursuing or adhering to a particular decision. The resultant consumer decision is not the result of reasoned decision-making on the basis of mandated information, but the result of the desire to avoid time, decision effort, or emotional costs. These costs may be ignored by a rational economic actor, but they influence biased consumers' ability to pursue or adhere to a particular choice.[190]

   Category 4 of autonomy violations is derived from the regulation of specific dark patterns mentioned in the DSA and the CRD.[191] In particular, two of the practices listed in DSA art. 25(3) are nagging and "making the procedure for terminating a service more difficult than subscribing to it."[192] One example of nagging in online choice architectures is the repeated requesting that a consumer consents to data processing with a pop-up that cannot be ignored and must be actioned even though consent has already been refused.[193] The pop-up request only disappears once a consumer consents to the data processing. The prohibition of dark patterns on online interfaces of online platforms should also apply to "making certain choices more difficult or time-consuming than others, making it unreasonably difficult to discontinue purchases or to sign out from a given online platform . . . or . . . default settings that are difficult to change."[194] These cases are also capable of falling under Category 4. Category 4 is also relevant for the design of privacy settings and cookie banners in online choice architectures. For instance, Category 4 applies to a cookie consent banner which requires one click to "accept all" optional cookies but has no "reject all" option and requires the user who intends to reject all optional cookies to individually set the toggle for, for example, fifty cookies on the "No" option.[195] Another example that is captured by Category 4 is an online choice architecture which effectively deters consumers from changing their (default) privacy preferences by hiding privacy settings behind multiple websites, tabs, and confusing settings.

---

[190] This Article focuses on frictions that affect biased consumers and their decision-making but that would not affect rational consumers. Consequently, the term "friction" is not employed in this Article as an indicator of rational consumer behavior in contrast to consumer behavioral biases. Thus, I deviate from Andrea La Nauze & Erica Myers, *Do Consumers Acquire Information Optimally? Experimental Evidence From Energy Efficiency* 1 (NBER Working Paper 31742, 2023), http://www.nber.org/papers/w31742 [https://perma.cc/LF7T-3PF5], who use the term "frictions" when referring to rational consumer behavior and the term "mental gaps" when referring to biased consumer behavior.

[191] *See* DSA, *supra* note 5, art. 25(3)(b) and (c); *id.* at recital 67; CRD, *supra* note 63, art. 16e(1)(b) and), (c).

[192] *See* DSA, *supra* note 5, art. 25(3)(b) and (c).

[193] *See* DSA, *supra* note 5, art. 25(3)(b). Mills et al., *supra* note 32, at 3 appear to argue that nagging is not an autonomy violation because it does not limit an individual's freedom of choice. This view neglects that autonomy also protects the independence of an individual's process of decision-making. *See id.* at art. 25(3)(b).

[194] DSA, *supra* note 5; *id.* at recital 67.

[195] This example is also a Category 5 autonomy violation. *See infra* Part III.B.5.

Specifying the autonomy violation that is targeted by the specific practices listed in DSA art. 25 for b2c relationships helps interpreting this provision.[196] For example, I contend that the DSA does not prohibit an online interface design which makes "the procedure for terminating a service more difficult than subscribing to it,"[197] unless it effectively deters consumers from terminating a service. The mere finding of an asymmetry in the level of difficulty between subscribing to and terminating a service does not determine an autonomy violation.[198] I am not aware of any autonomy theory which would posit an autonomy violation simply because the cancellation procedure requires two simple clicks on a website, whereas the procedure to sign up to the service requires only one click. In this example, the online choice architecture does not deter consumers from terminating a service. Consumers' ability to make a model self-determined decision is not violated. The decisions to subscribe to and cancel a service are two separate decisions, which often occur at completely different points in time and which warrant two separate assessments. Autonomy theory does not demand that it must be as easy to cancel a subscription as it was to sign up to it. This interpretation of DSA art. 25 does not require a reading down of the statutory words. That is because the introductory statutory words of DSA art. 25(3) suggest that the listed practices are not prohibited in all circumstances. Category 4 of autonomy violations determines in which circumstances the practices listed in DSA art. 25(3)(b) and (c) amount to a violation of autonomy.[199]

Category 4 of autonomy violations is particularly relevant for "subscription traps," that is the design of an online interface in such a way that it effectively deters consumers from unsubscribing to a service by creating unreasonable time, decision effort, or emotional costs. For example, consumers had to go through multiple cumbersome steps to cancel their Amazon Prime subscription before Amazon changed its cancellation practice in the EU in 2022 after a dialogue with the European Commission. Such steps involved scrolling through multiple pages that contained unclear button labels, warnings that deterred users from cancelling, skewed wording, confusing choices, and complicated navigation menus.[200] As pointed out in the previous paragraph, this cancellation procedure is not an autonomy violation simply because an asymmetry exists in the level of difficulty between subscribing to and cancelling Amazon Prime. What matters is whether consumers face unreasonable decision-making costs when cancelling a subscription online, which effectively deters consumers from pursuing their decision to cancel the subscription. If the process to subscribe to a service is easier than the procedure for cancelling a service, the ease of the subscription process may indicate how easy the cancellation process could have been. This leads to the question of whether the time, decision effort, or emotional costs involved in the actual cancellation process are unreasonable compared to the decision costs involved in the hypothetical cancellation process.

---

[196] This also applies to CRD, *supra* note 63, art. 16e due to the largely identical terms and practices listed in both provisions.

[197] DSA, *supra* note 5, art. 25(3)(c).

[198] This position is supported by the legislative history of DSA, *supra* note 5, art. 25. The original proposal by the European Parliament intended to prohibit "making the procedure of terminating a service significantly more cumbersome than signing up to it" because this practice "distort[s] or impair[s] recipients of services' ability to make a free, autonomous and informed decision or choice." (European Parliament, *Digital Services Act* (Jan. 20, 2022), https://www.europarl.europa.eu/RegData/seance_pleniere/textes_adoptes/definitif/2022/01-20/0014/P9_TA(2022)0014_EN.pdf [https://perma.cc/QKK8-F432]). Hence, the European Parliament identified a violation of autonomy not because the procedure for terminating a service was more difficult than signung up to it but because this procedure was *significantly* more difficult than signing up to it. This initial formulation by the European Parliament is now contained in DSA recital 67 ("making the procedure of cancelling a service significantly more cumbersome than signing up to it").

[199] Note that a prohibition of the listed practices is only justified if the autonomy violation is "material" (see the wording of DSA, *supra* note 5, art. 25(1)). The materiality condition in DSA art. 25(1) incorporates a weighing of multiple factors like the extent of the autonomy harm, the extent of the online platform's gain and other possible consumer harms which stand outside the autonomy violation. *See* Brenncke, *supra* note 99, at 17.

[200] *See, in detail*, European Commisson, *Consumer Protection: Amazon Prime Changes its Cancellation Practices to Comply With EU Consumer Rules* (July 1, 2022), https://ec.europa.eu/commission/presscorner/detail/en/ip_22_4186; Forbrukerrådet, *You Can Log*

Category 4 of autonomy violations also has the potential to rein in drip pricing practices. According to the DSA, providers of online platforms should be prohibited from "making it unreasonably difficult to discontinue purchases."[201] When the total cost of a product or service only becomes visible after consumers have invested a significant amount of time and effort in researching, analysing, making, and pursuing a particular choice option, biased consumers may be unwilling to waste these sunk costs. Biased consumers may complete the purchasing process even if they had not started this process had they known about the total price at the beginning of the purchasing process. This shows that the time, decision effort, or emotional costs that a consumer invests in the purchasing process can make it difficult for biased consumers to pursue the decision to discontinue the purchase once the total price is revealed. Whether the additional costs are unreasonable is an evaluative decision. Since mandatory fees must be included in the total price,[202] it is possible for the business to reveal those fees at the beginning of the purchasing process and incorporate them in the headline price.[203] It is difficult to see a reason for revealing mandatory fees during or at the end of the purchasing process, apart from taking advantage of behavioral factors. This reasoning supports the proposition that the dripping of mandatory fees creates unreasonable costs and constitutes a Category 4 autonomy violation.

Before moving on to the next Category of autonomy violations, it is worth pointing out that Category 4 of autonomy violations must be distinguished from two other types of friction in online choice architectures. First, the unreasonableness criterion excludes friction in online choice architectures which adds time, decision effort, or emotional costs, but which does not effectively deter biased consumers from pursuing or adhering to a particular decision. Such choice architecture may be a nuisance, but it does not violate consumers' ability to make a model self-determined decision. Second, Category 4 must also be distinguished from "positive friction", which refers to choice architecture that is able to trigger or increase reflective decision-making and support a model autonomous decision.[204] Even though difficult issues of demarcation exist between positive and negative friction,[205] such questions fall outside the scope of this Article because dark patterns in private online choice architectures do not trigger or increase reflective decision-making on the basis of mandated information. The unreasonableness criterion in Category 4 is also able to capture the distinction between negative and positive friction.

### 5. *Category 5: Non-neutral Presentation of Choice Options*

*Dark patterns in online choice architectures violate consumers' ability to make a model self-determined decision if they present choice options in a non-neutral manner when asking consumers to select between different choice options.*

Category 5 specifies the autonomy violation addressed by DSA art. 25(3)(a), which mentions the specific practice of "giving more prominence to certain choices when asking the recipient of the service for a decision," for example through "visual, auditory, or other components."[206] The CRD enumerates a nearly identical practice.[207] The Data Act also addresses Category 5 autonomy violations when stipulating that third parties and data holders are prohibited from making the exercise of certain user choices "unduly difficult, including by offering choices to the user in a non-neutral manner."[208] These provisions in the Data Act, the DSA, and the CRD contain the legislative generalization that an autonomy violation is present in b2c relationships if the online choice architecture presents choice options in a non-neutral manner when asking consumers to select between different choice options. This legislative assessment is compatible with autonomy theory, but requires a differentiation.

When the consumer reflects about the non-neutral design, Category 5 can be a specific case of a Category 4 autonomy violation if the non-neutral design effectively deters biased consumers from pursuing one of the choice options. The consumer favors one choice option over the other due to the desire to avoid time, decision effort, or emotional costs. More commonly, Category 5 captures influences on consumer decision-making at the non-reflective level. Part II.C of this Article already explained that (i) presentation effects like salience trigger consumer biases when they affect consumer decision-making and (ii) these effects typically work at the non-reflective level. For example, consumers do not usually reflect on whether or to what extent the prominent color or size of an agree button compared to a reject button influences their decision-making. Biased consumers may not even realize the existence of a reject option if the agree option is presented in an overly prominent manner. If such influences on consumers' non-reflective decision-making are successful in steering consumers towards one of the choice options, consumers are not aware of the factors that can explain their decision.

---

[204] *Cf.* Lucy Hayes et al., *Beyond Disclosure for High-Risk Investments: Slow Down and Think* (Feb. 2, 2022), https://www.fca.org.uk/publications/research/beyond-disclosure-high-risk-investments-slow-down-and-think [https://perma.cc/7VY7-F3UR] (defining positive frictions as "processes designed to slow people down and make them consider their actions more carefully", but adopting a wide understanding of the term, since positive frictions "might alter people's behavior through many channels, not just by inducing reflection and in turn in a fuller comprehension of relevant information"). *See also* Brett Frischmann & Susan Benesch, *Friction-in-design Regulation as 21st Century Time, Place, and Manner Restriction*, 25 YALE J.L. & TECH. 376, 388–91 (2023) (highlighting possible benefits of design interventions which create frictions in consumer decision-making and enable consumers to stop and reflect).
[205] *See, e.g.*, Alemanno & Sibony, *supra* note 107, at 331–32.
[206] DSA, *supra* note 5, recital 67.
[207] *See* CRD, *supra* note 63, art. 16e(1)(a).
[208] Data Act, *supra* note 71, arts. 4(4) and 6(2)(a).

The main challenge in establishing a Category 5 autonomy violation lies in accurately assessing the instances when choice options are presented in a "non-neutral" manner. I contend that the presentation of choice options only qualifies as "non-neutral" if it effectively steers biased consumers towards one of the choice options. Multiple reasons support this view. First, based on a purely empirical perspective, it could be argued that choice options can never be presented in a completely neutral manner.[209] Every choice architecture is non-neutral. However, the DSA and the Data Act imply that choice options can be presented in a neutral manner.[210] Second, the Data Act prohibits data holders from making the exercise of certain choices by the user "unduly difficult, including by offering choices to the user in a non-neutral manner."[211] Trivial asymmetrical presentations of choice options do not make the exercise of these choices unduly difficult. Hence, trivial asymmetrical presentations in online choice architectures which do not effectively steer biased consumers towards one of the choice options can be considered as normatively neutral.[212] Third, not every successful external influence on consumers' non-reflective decision-making processes violates consumers' ability to make a model self-determined decision.[213] An autonomy violation can be inferred if the asymmetrical presentation of choice options effectively steers biased consumers towards one of the choice options. This understanding of Category 5 of autonomy violations is also compatible with DSA art. 25(3)(a). That is because the introductory statutory words of DSA art. 25(3) suggest that the listed practices are not prohibited in all circumstances. Category 5 of autonomy violations determines in which circumstances the practice listed in DSA art. 25(3)(a) amounts to a violation of autonomy.[214]

The following examples are Category 5 autonomy violations if they effectively steer biased consumers towards one of the choice options: (i) the design of a cookie consent banner that gives more prominence to the "Agree to all" button compared to the "Reject all" button; (ii) the design of a cookie consent banner that allows consumers to accept a privacy policy with one click (for example, "Accept and continue" button) but does not allow consumers to deny the privacy policy with one click (for example, "Other options" button); (iii) the pre-selection of one choice option with a pre-ticked box; and (iv) the choice given to a consumer between "No, stay a member" and "Yes, cancel membership," when the business gives more prominence to the first choice option by using a larger font size, a colored font, or by presenting the option within a prominent box design.

---

[209] *See* Jacob L. Orquin et al., *Visual Biases in Decision Making*, APPLIED ECON. PERSP. & POL'Y 523 (2018).

[210] *Cf.* DSA, *supra* note 5, recital 67; Data Act, *supra* note 71, arts. 4(4) and 6(2)(a).

[211] Data Act, *supra* note 71, art. 4(4).

[212] Similar considerations apply to DSA, *supra* note 5, art. 25(3)(a) and CRD, *supra* note 63, art. 16e(1)(a).

[213] *See supra* Part III.A.

[214] *See infra* note 199. This interpretation of DSA, *supra* note 5, art. 25(1), (3)(a) also applies analogously to CRD, *supra* note 63, art. 16e due to the largely identical terms and practices listed in both provisions.

### 6. *Category 6: Manipulation*

*Dark patterns in online choice architectures violate consumers' ability to make a model self-determined decision if they manipulate consumers.*

Category 6 of autonomy violations is derived from DSA art. 25(1), which prohibits providers of online platforms to use online interface designs that manipulate consumers. Equivalent prohibitions exist in the CRD and the Data Act.[215] The wording of DSA art. 25(1) implies that manipulation is a material distortion or impairment of consumers' ability to make free and informed decisions. Hence, a manipulative interface design on online platforms is an autonomy violation that warrants legal intervention. Unfortunately, both the recitals and the legislative history of the DSA fail to specify the term manipulation. This assessment similarly extends to the CRD and the Data Act. This is a significant omission because the meaning of the term is far from clear. There is considerable disagreement in the literature about the necessary and sufficient elements of manipulation.[216]

Irrespective of these controversies, there appears to be a large agreement in scholarship that manipulators wrong their victims as autonomous agents.[217] For example, Susser et al. see the harm and wrong-making feature of manipulation as a violation of personal autonomy.[218] Their influential account of manipulation defines manipulating someone as "intentionally and covertly influencing their decision-making, by targeting and exploiting their decision-making vulnerabilities," such as cognitive biases.[219] If, as Susser et al. claim, "manipulation functions by exploiting the manipulee's cognitive (or affective) weaknesses and vulnerabilities,"[220] exploitation is a key element of manipulation. Exploitation is generally considered to be a moralized concept in philosophical discourse,[221] and multiple normative theories can explain the wrongness of exploitation.[222] In the autonomy theory of exploitation, for example, exploitation is wrong because the exploiter treats the exploitee in a way that violates the latter's personal autonomy.[223] This theory of exploitation aligns well with Susser et al's definition of manipulation, since the wrong-making feature of manipulation lies in a violation of autonomy.

Importantly, autonomy theory serves as the yardstick for determining when the exploitation of consumer biases is wrong due to the violation of autonomy. The violation of the autonomy benchmark has already been specified in Categories 1 to 5 of autonomy violations, for the purposes of EU consumer law governing the regulation of dark patterns in online choice architectures.[224] This essentially means that Categories 1 to 5 of autonomy violations specify the autonomy violation inherent in the term manipulation. This reference to other categories of autonomy violations makes the term manipulation legally operational and provides legal certainty. Even though this interpretation of manipulation renders Category 6 superfluous in terms of establishing an autonomy violation, the Category remains relevant for assessing whether the autonomy violation warrants legal intervention. That is because the term manipulation further qualifies the autonomy violation from Categories 1 to 5. First, there are additional exploitation criteria such as the exploiter's gain.[225] Second, the influence on consumer decision-making must be intentional and covert. Intentionality and hiddenness are often recognized as two necessary elements of manipulation.[226] These qualifying factors explain why the inherent autonomy violation within manipulation is deemed "material"[227] and warrants prohibition when it occurs in online interface designs of online platforms.

---

[215] *See* CRD, *supra* note 63, art. 16e(1); Data Act, *supra* note 71, art. 6(2)(a).

[216] *See, e.g.*, Marcello Ienca, *On Artificial Intelligence and Manipulation*, 42 TOPOI 833–842 (2023) (discussing different accounts of manipulation in the literature).

[217] *See* Mills, *supra* note 154, at 223.

[218] *See* Susser et al., *supra* note 107, at 35.

[219] Daniel Susser et al., *Technology, Autonomy, and Manipulation*, 8 INTERNET POL'Y REV. 1, 4 (2019).

[220] Susser et al., *supra* note 107, at 3.

[221] *See, e.g.*, Joel Feinberg, *Noncoercive Exploitation*, in PATERNALISM 201, 202 (Rolf Sartorius ed., 1983); Robert E. Goodin, *Exploiting a Situation and Exploiting a Person*, in MODERN THEORIES OF EXPLOITATION 166, 173 (Andrew Reeve ed., 1987); ALAN WERTHEIMER, EXPLOITATION 6 (1996).

[222] *See* Brenncke, *supra* note 99, at 21–30; MATHIAS RISSE & GABRIEL WOLLNER, ON TRADE JUSTICE: A PHILOSOPHICAL PLEA FOR A NEW GLOBAL DEAL 81–85 (2019); Nicholas Vrousalis, *Exploitation: A Primer*, 13 PHIL. COMPASS 1, 2–11 (2018) (all discussing different classes of theories that determine the moral wrong of exploitation).

[223] *See, e.g.*, Brenncke, *supra* note 99, at 26; RICK BICKWOOD, EXPLOITATIVE CONTRACTS chapter 3, 203, 225 (2003).

## C. CONCLUSION

The six categories of autonomy violations function as a normative classification for dark patterns. This categorization addresses the limitations present in current taxonomies of dark patterns found in the literature, which are not adequately suited for the requirements of EU consumer law. Most classifications of dark patterns in the non-legal literature are descriptive and based on (i) technical characteristics of dark patterns like "trick questions," "roach motel," "sneak into basket," "fake countdown timers," and "disguised ads" or (ii) the technique used by dark patterns such as nagging, obstructing, sneaking, or forced action.[228]

Normative classifications of dark patterns that adopt an autonomy framework exist, but are rare. For example, Shuja and Kumar assess the ethical concerns raised by dark patterns, and they use personal autonomy as a normative benchmark to identify these concerns.[229] They work with a conception of personal autonomy that is derived from the philosophical literature on autonomy and the literature on nudging and dark patterns. Their taxonomy is insufficiently aligned with the specific conception of autonomy that underlies EU consumer law. Leiser and Yang develop a normative taxonomy of dark patterns which is consistent with the legislative structure of the UCPD.[230] They do not address the crucial issue of how extensively the average consumer model in the UCPD limits the Directive's ability to protect biased consumers from the influence of dark patterns. Their taxonomy is also insufficiently aligned with the provisions expressly regulating dark patterns in EU consumer law, which were analyzed in Part I of this Article.

---

[224] This includes Category 2, since deception is "an important tool in the manipulator's toolkit" (Susser et al., *supra* note 107, at 21).

[225] *See* Brenncke, *supra* note 99, at 5; Feinberg, *supra* note 221, at 203; Goodin, *supra* note 221, at 173; WERTHEIMER, *supra* note 221, at 17.

[226] *See, e.g.*, Ienca, *supra* note 216, at 837–38; Susser et al., *supra* note 107, at 26.

[227] DSA, *supra* note 5, art. 25(1).

[228] *See* Gray et al., *supra* note 28, at 10–11; Leiser & Yang, *supra* note 44, at 2–14; Mathur et al., *supra* note 10; Mills et al., *supra* note 32, at 5–8; Marie Potel-Saville et al., *From Dark Patterns to Fair Patterns? Usable Taxonomy to Contribute Solving the Issue With Countermeasures* 8–10 (2023),
https://www.researchgate.net/publication/371314839_From_Dark_Patterns_to_Fair_Patterns_Usable_Taxonomy_to_Contribute_Solving_the_Issue_with_Countermeasures (all discussing different taxonomies of dark patterns in the literature).

[229] *See* Sanju Ahuja & Jyoti Kumar, *Conceptualizations of User Autonomy Within the Normative Evaluation of Dark Patterns*, 24:52 ETHICS & INFO. TECH. 1–18 (2022).

In terms of their granularity, the six categories of autonomy violations are placed between detailed rules that target specific cases of dark patterns and a vague and unpredictable general clause.[231] Even though the application of these categories to individual cases may involve moral and policy considerations, the categories avoid the vagueness of a general clause on the one hand and the risk that designers of online choice architectures can circumvent specific lists of regulated practices by inventing new cases of dark patterns on the other hand. For the purposes of specifying the prohibition of dark patterns in the DSA, for example, the six categories of autonomy violations provide a middle layer between the general clause, a material distortion, or impairment of the ability to make free and informed decisions,[232] and the specifically enumerated practices in DSA art. 25(3).

CONCLUSION

Dark patterns are an increasingly prevalent issue in online choice architectures. Regulators around the world are intensifying their efforts to effectively regulate these practices. This is normatively challenging for multiple reasons. First, dark patterns often operate in the grey zone between legitimate persuasion techniques and clearly illegitimate methods of influencing consumer behavior such as coercion and deception. Second, dark patterns in online choice architectures work because they exploit consumer behavioral biases. This is a challenge for traditional EU consumer legislation like the UCPD, which is said to adopt the consumer image of a rational economic actor. Third, the notion of protecting consumer autonomy as a normative rationale for assessing and regulating dark patterns is under-researched. The existing literature has not yet produced a well-developed autonomy framework for regulating dark patterns in the EU. This Article has addressed these normative challenges. Its principle theses can be summarized as follows:

1.  Multiple legislative acts in the EU either expressly protect consumers from dark pattern practices in online choice architectures or, equivalently, expressly protect consumers from commercial practices in online choice architectures that exploit consumer behavioral biases. The key provisions are DSA art. 25, CRD art. 16e, Data Act arts. 4(4) and 6(2)(a), CRD art. 22, CCD II art. 15, and the Key Information Document in the PRIIPs Regulation. These provisions (also) protect biased consumers. The normative theory that these provisions adopt to assess dark patterns is autonomy. Regulating dark patterns in EU law means regulating for autonomy.

---

[230] *See* Leiser & Yang, *supra* note 44.
[231] *Cf.* Osmola, *supra* note 34 (discussing the rules versus standard paradigm specifically for the issue of regulating dark patterns).
[232] *See* DSA, *supra* note 5, art. 25(1).

2.  A specific meaning of autonomous decision-making can be extracted from the information paradigm in EU consumer law. The information paradigm does not assume that consumers are rational economic actors. Instead, it can have an autonomy-based justification. The information paradigm enables consumers to make model self-determined decisions, and the provisions regulating dark patterns protect consumers' ability to make model self-determined decisions. A model self-determined decision is a decision based on mandated information, which the consumer correctly understands and which generates reasons in the decision-making process. Consumers are aware of these reasons that can explain and justify their decision.

3.  Drawing upon the existing provisions regulating dark patterns in EU consumer law, a taxonomy encompassing six categories of autonomy violations can be developed. These categories serve as a normative classification for dark patterns in b2c relationships. These categories not only uncover and make explicit the autonomy violations addressed by existing EU provisions governing dark patterns but also offer policymakers a framework when deliberating the regulation of other instances of dark patterns.

4.  Dark patterns in online choice architectures violate consumers' ability to make a model self-determined decision if they:
    A.  Influence consumers' decision-making in such a way that consumers ignore or misunderstand mandated information;
    B.  Cause consumers to hold false beliefs that form the foundation of consumers' decision-making;
    C.  Cause consumers to enter into a specific contractual agreement with a business without reflection about its substance and content;
    D.  Create unreasonable time, decision effort, or emotional costs for pursuing or adhering to a particular decision;
    E.  Present choice options in a non-neutral manner when asking consumers to select between different choice options; or
    F.  Manipulate consumers.